\documentclass[conf]{new-aiaa}
\usepackage[utf8]{inputenc}

\usepackage{graphicx}
\usepackage{amsmath}
\usepackage[version=4]{mhchem}
\usepackage{siunitx}
\usepackage{longtable,tabularx}
\usepackage{subfigure}% subcaptions for subfigures
\usepackage{float}
\usepackage{overpic}

\usepackage{xcolor}					%definir novas cores

\title{Comparison of Shock-Boundary Layer Interactions in \\ Adiabatic and Isothermal Supersonic Turbine Cascades}

\author{Hugo F. S. Lui\footnote{Graduate Student, School of Mechanical Engineering, E-mail: hugo.slui@gmail.com.}, Tulio R. Ricciardi \footnote{Postdoctoral Research Associate, National Center for Supercomputing Applications, University of Illinois Urbana-Champaign, E-mail: tricci@illinois.edu.} and William R. Wolf\footnote{Associate Professor, School of Mechanical Engineering, Department of Energy, E-mail: wolf@fem.unicamp.br.}}
\affil{University of Campinas, UNICAMP, Campinas, SP, 13083-860, Brazil}
\author{Carlos A. J. B. Junior\footnote{Research Engineer, Laboratoire de Dynamique de Fluides (Dynfluid), E-mail:  junior.junqueira@ensam.eu.}}
\affil{École Nationale Supérieure d'Arts et Métiers, ENSAM, Paris, 75013, France}

\begin{document}

\maketitle

\begin{abstract}
Wall-resolved large eddy simulations are employed to investigate the shock-boundary layer interactions (SBLIs) in a supersonic turbine cascade. An analysis of the suction side separation bubbles forming due to the SBLIs is presented for adiabatic and isothermal (cooled) walls. Flow snapshots indicate that the separation bubble contracts and expands in a similar fashion for both thermal boundary conditions. However, the skin-friction coefficient distributions reveal a downstream displacement of the separation region when cooling is applied. The separation bubble is also smaller for this setup compared to the adiabatic one. A steeper pressure rise is observed for the isothermal wall downstream of the incident oblique shock, and this occurs because the incident shock wave gets closer to the blade surface when cooling is applied. The Reynolds stresses are computed to investigate the effects of wall temperature on the turbulence activity. While the levels of the tangential stresses are similar for the cases analyzed, those for the wall-normal component are higher for the cooled wall.

\end{abstract}

\section{Introduction}
\lettrine{S}{upersonic} fluid machinery are applied in high-speed propulsion and power generation systems due to their high power density \cite{PANIAGUA201465}. In supersonic turbines, inlet shock waves are formed and interact with the boundary layers of neighboring blades. The shock-boundary layer interactions (SBLIs) can increase the aerodynamic drag due to flow separation and induce higher heat transfer rates to the blade surface. They can also be a source of flow unsteadiness, where multiple frequencies are excited due to motion of the incident and reflected shock waves, breathing of the separation bubble, besides the incoming turbulent boundary layer. Typically, the shock wave motion leads to strong pressure fluctuations that can compromise the system's structural integrity \cite{babinsky_harvey_2011, GAITONDE2015,Lui2021,lui2022}. 

Most studies of SBLIs have considered adiabatic wall conditions and, thus, the effects of surface heat transfer are not fully explored. Schülein \cite{Erich2006} used non-intrusive techniques to perform heat transfer and skin-friction measurements in the impingement of an incident oblique shock wave on a flat plate with isothermal wall conditions. Their results show that within the separation region, the heat flux increases in the streamwise direction, while the skin-friction decreases. Jaunet et al. \cite{Jaunet2014} investigated experimentally the impact of the wall temperature on a shock-induced boundary layer separation. They observed that the interaction length considerably increases when the wall temperature is raised. Bernardini et al. \cite{Bernardini2016} and Volpiani et al. \cite{Volpiani2018} carried out direct numerical simulations (DNS) to investigate the wall temperature effects on the physics of SBLIs. Results revealed that wall cooling significantly reduces the size of the separation bubble and interaction scales, while the opposite behavior is noticed in the case of wall heating.

In the present work, a high-order overset compressible large eddy simulation (LES) methodology is employed to investigate the flow in a supersonic turbine cascade with two different wall thermal boundary conditions. These include an adiabatic and a cooled walls, where the wall to inlet temperature ratio is set as $T_w/T_\infty = 0.75$. First, the numerical methodology is described including the grid details and flow configurations. Spanwise and time averaged pressure and skin-friction coefficients, as well as the mean flow fields, are presented to assess the effect of cooling on the size and form of the separation bubble. Then, flow snapshots are analyzed to investigate the features of the separation bubbles and the shear layer dynamics at different instants of the SBLI. Finally, the effects of the wall thermal boundary conditions on the turbulence activity are analyzed by assessing the Reynolds stress distributions. 

\section{Numerical Methodology} 

%\subsection{Large Eddy Simulations}

The present wall-resolved large eddy simulations solve the compressible Navier-Stokes equations in a curvilinear coordinate system. The fluid is assumed to be a calorically perfect gas, where the molecular viscosity $\mu$ is considered to depend on the local temperature through the nondimensional Sutherland's law. The spatial discretization of the governing equations is performed using a sixth-order accurate finite-difference compact scheme \citep{Nagarajan2003} implemented on a staggered grid. A sixth-order compact interpolation scheme is also used to obtain flow quantities on the different nodes of the staggered grid configuration.

Two grids are employed in the present simulations: one is a body-fitted O-grid block which surrounds the airfoil and the other is an H-grid block used to enforce the pitchwise periodicity of the cascade. In the O-grid, the time integration of the equations is carried out by the implicit second-order scheme of \citet{Beam1978}. This method overcomes the stiffness problem arising from the wall-resolving boundary layer mesh. In the background H-grid block, a third-order Runge-Kutta scheme is used for time advancement of the governing equations. A fourth-order Hermite interpolation scheme \citep{Delfis2001,Bhaskaran} is used to exchange information between grid blocks in an overlapping zone.
Further details about the numerical procedure can be found in \citep{Bhaskaran}.

Due to the non-dissipative characteristics of the compact finite-difference schemes, numerical instabilities may arise from mesh non-curvature, interpolation between the overset grids, and boundary conditions. To preserve stability of the numerical simulations, the high wavenumber compact filter presented by \citet{Lele1992} is applied in flow regions far away from solid boundaries at each time step. A shock capturing scheme is also employed to capture the shock waves forming in the present flows. In order to introduce minimal numerical dissipation in the vicinity of the shocks, without damping the small scales of turbulence, the localized artificial diffusivity (LAD) method \citep{Cook2007} is employed to compute the artificial bulk viscosity and thermal conductivity. The approach LAD-D2-0 proposed by \citet{Kawai2010} is employed here with no artificial shear viscosity. In order to transition the boundary layers, we apply a body forcing on the RHS of the Navier-Stokes equations, as described by \citet{Sansica}. Here, an unsteady actuation with a random spanwise treatment is assumed and the amplitude of the disturbances are chosen experimentally in order to guarantee a bypass transition with minimal flow disturbance. More details of the numerical procedure can be found in \citep{lui2022}.

\section{Flow and Mesh Configurations}

\begin{figure}[H]
	\begin{overpic}[trim = 17mm 10mm 15mm 10mm, clip,width=0.99\textwidth]{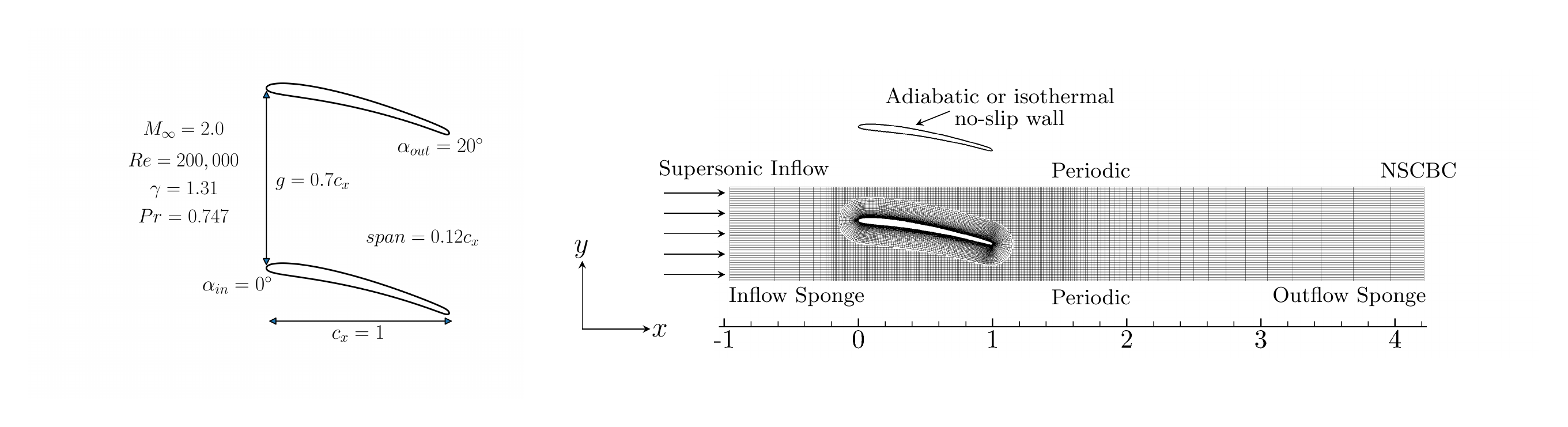}
		\put(0,20){(a)}
		\put(32,20){(b)}		
	\end{overpic} 
	\caption{Schematics of (a) flow configuration and geometrical parameters, and (b) computational domain skipping every $5$ grid points.}
	\label{fig:schematic}
\end{figure}

This section shows details of the flow configuration studied and describes the computational grid used in the LES calculations. Figure \ref{fig:schematic} (a) presents the geometrical parameters and flow conditions. The inlet Mach number is set as $M$ = 2.0 and the Reynolds number based on the inlet velocity $U_{\infty}$ and axial blade chord is $Re$ = 200,000. The ratio of specific heats is chosen as $\gamma = 1.31$, the Prandlt number is $Pr = 0.747$ and the ratio of the Sutherland constant over inlet temperature is set as $S_{\mu}/T_{\infty}$ = 0.07182. These conditions are chosen based on previous studies \cite{LIU2019,Lui2021, lui2022}.

Figure \ref{fig:schematic} (b) displays a schematic of the overset grid employed in the LES along with the implemented boundary conditions. The O-grid block has $1200 \times 280 \times 144$ points and is embedded in the background Cartesian grid block of size $960 \times 280 \times 72$. Therefore, the grid has approximately $68,000,000$ points. Depending on the case, adiabatic or isothermal boundary conditions are applied along the blade surface. For the latter, the wall to inlet temperature ratio is $T_w/T_{\infty}=0.75$, representing a cooled wall. Supersonic inflow boundary conditions are used to set the inlet conditions. For the outflow, a boundary condition based on the Navier-Stokes characteristic boundary condition (NSCBC) \citep{Poinsot1992} is employed. A damping sponge is also applied near the inflow and outflow boundaries to minimize reflections of disturbances \citep{ISRAELI1981,Nagarajan2003}. Periodic boundary conditions are used in the $y$-direction of the background grid, according to Fig.\ \ref{fig:schematic} (a), in order to simulate a linear cascade of blades and periodic boundary conditions are also applied in the spanwise direction, to enforce a statistically homogeneous flow along the span.

%Figure \ref{fig:schematic} shows a schematic of the overset grid employed in the LES along with the implemented boundary conditions. The computational grid is composed by a body-fitted O-grid block of size $1280 \times 300 \times 144$ which is embedded in a background Cartesian grid block of size $960 \times 280 \times 72$. Therefore, the LES grid has approximately $75\times10^6$ points. 

For the adiabatic wall case, the grid resolution in terms of wall units is kept in the range given by $6<\Delta s^+<25$, $0.1<\Delta n^+<0.3$, and $3<\Delta z^+<9$, where $s$, $n$ and $z$ represent the streamwise, wall-normal and spanwise flow coordinates. For the isothermal wall simulation, the near-wall grid spacing ranges from $15<\Delta s^+<60$, $0.2<\Delta n^+<0.6$, and $6<\Delta z^+<19$. These numbers are computed for regions where the boundary layers are fully developed and in equilibrium, away from the tripping and recirculation regions. It is worthwhile to mention that the same computational grid is used for both cases, but higher values in terms of wall units are obtained for the isothermal wall case due to a inherent reduction of the viscous length scales caused by cooling.

The simulation is initialized with a uniform flow and statistics are computed after the initial transients are discarded.
%, where a tangent-hyperbolic velocity profile is applied near the blade surface in order to satisfy the no-slip boundary condition and to smooth out initial velocity gradients.
In the simulations, a variable time step is computed based on an inviscid CFL parameter of 0.8. The body-force tripping is applied at $0.22 < x/c_{ax} < 0.27 $ for the suction side, and at $0.10 < x/c_{ax} < 0.15 $ for the pressure side. The wall normal height of the body-foce region is $\delta = 0.001c_{ax}$ and the actuation changes every $\Delta t \approx$ 0.003 in a spanwise-random fashion. 

\section{Results}

\begin{figure}[H]
	\centering	
	\begin{overpic}[trim = 1mm 1mm 1mm 1mm, clip,width=0.48\textwidth]{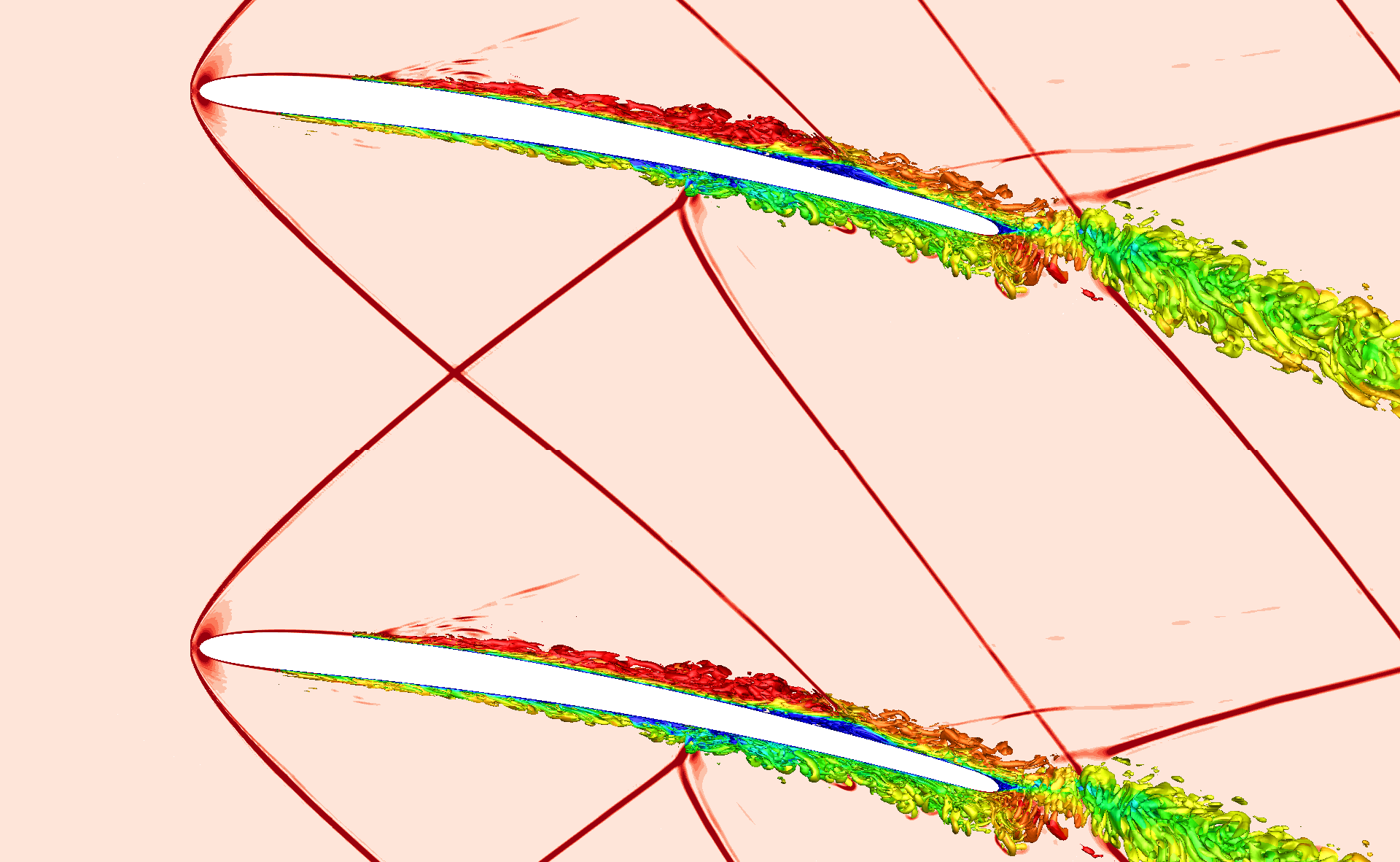}
		\put(1,58){(a)}
	\end{overpic} 
	\begin{overpic}[trim = 1mm 1mm 1mm 1mm, clip,width=0.48\textwidth]{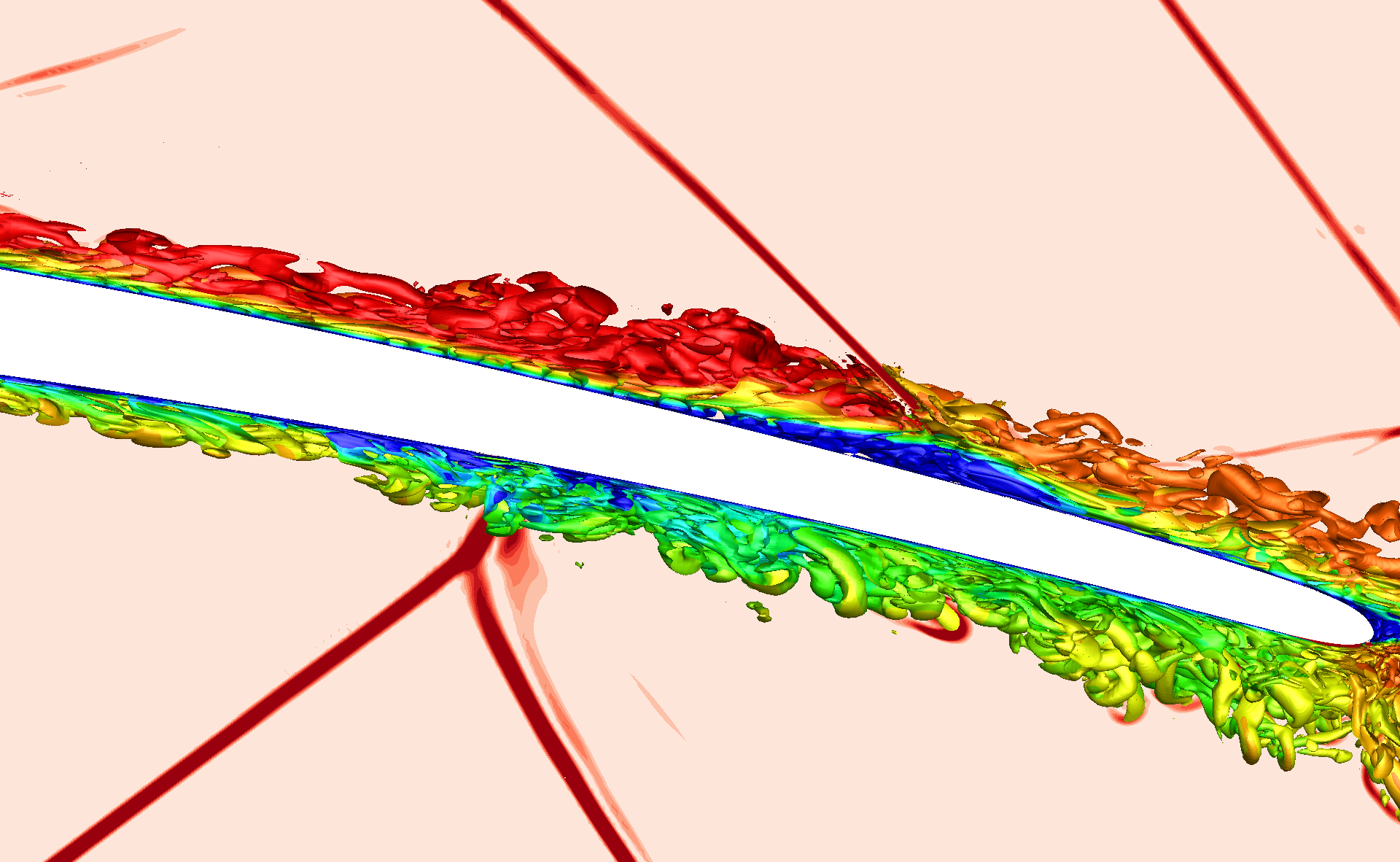}
		\put(1,58){(b)}
	\end{overpic} \\
	\vspace{2mm}
	\begin{overpic}[trim = 1mm 1mm 1mm 1mm, clip,width=0.48\textwidth]{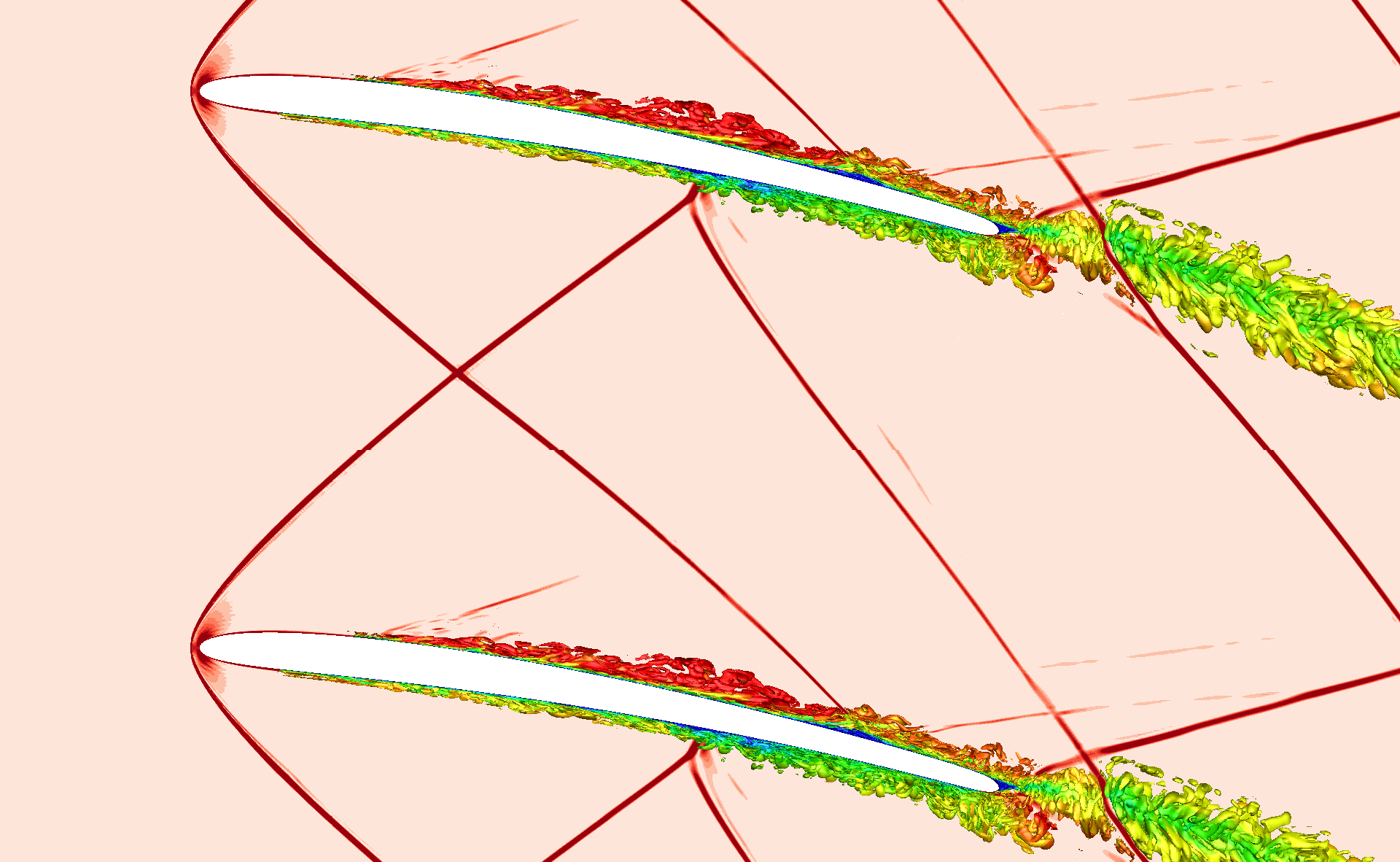}
		\put(1,58){(c)}
	\end{overpic} 
	\begin{overpic}[trim = 1mm 1mm 1mm 1mm, clip,width=0.48\textwidth]{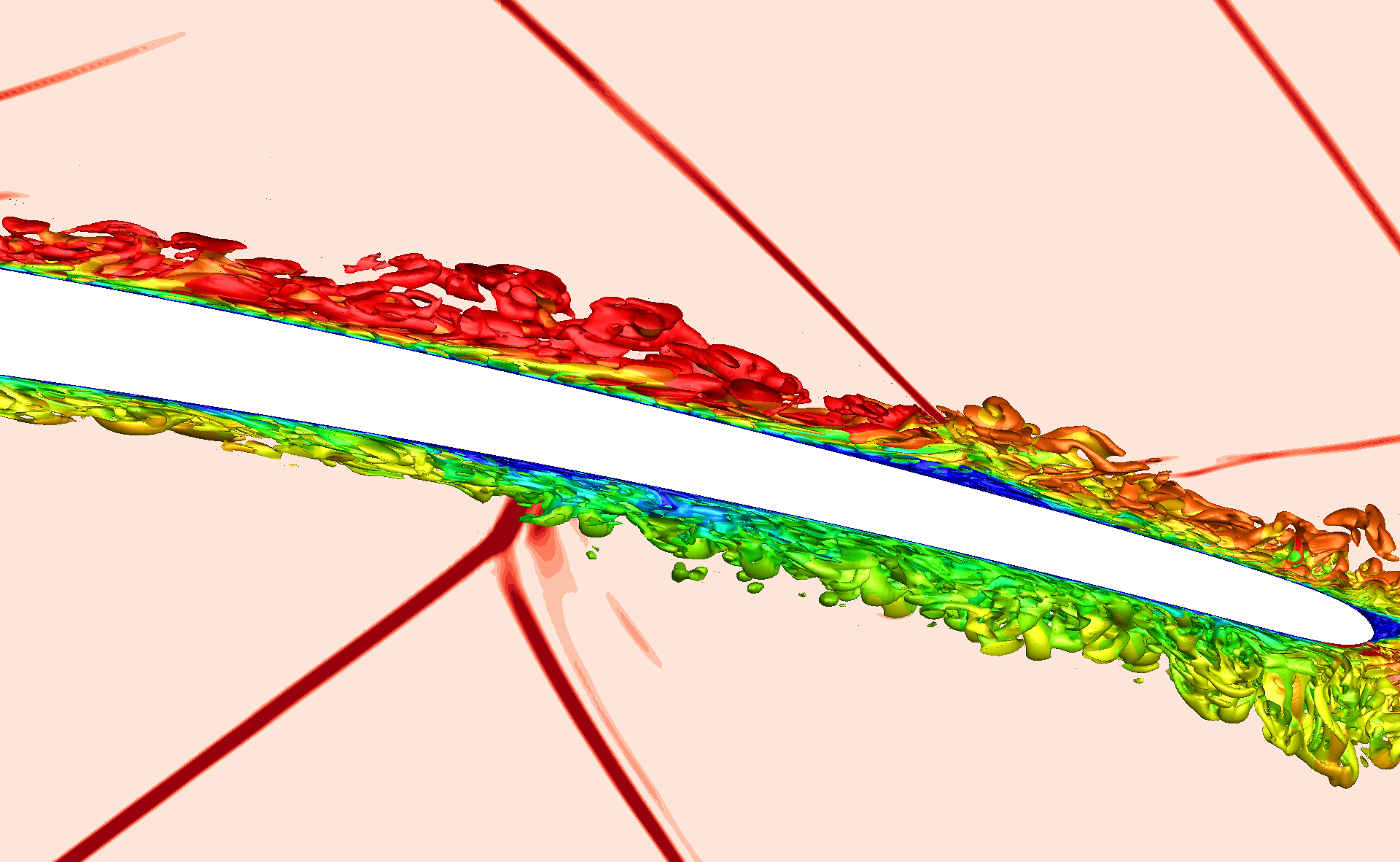}
		\put(1,58){(d)}
	\end{overpic}
	\caption{Iso-surfaces of $Q$-criterion colored by $u$-velocity component for the adiabatic (top) and cooled (bottom) wall cases. The background plane displays the shock waves by visualizing the density gradient magnitude $|\nabla \rho|$.}
	\label{fig:q_criterion}
\end{figure}

This section presents results obtained by the LES computed for adiabatic and isothermal (cooled) wall boundary conditions. Flow quantities are collected for 4 flow through times, based on the inlet velocity and blade axial chord. Figure \ref{fig:q_criterion} shows iso-surfaces of $Q$-criterion colored by the $u$-velocity component together with a background view of density gradient magnitude, $|\nabla \rho|$. The top and bottom rows present results for the adiabatic and cooled wall cases, respectively.

In Figs. \ref{fig:q_criterion} (a) and (c), we can observe the complex shock structure across the turbine passage. The detached oblique shock waves generated at the airfoil leading edges interact with the boundary layers of the neighboring blades and are reflected across the cascade. On the pressure side, the incident shock wave becomes normal to the wall and, then, a Mach reflection is formed, while an oblique shock reflection is generated on the suction side. To 
highlight the effect of cooling on the SBLI, a detailed view of the flow field can be seen in Figs. \ref{fig:q_criterion} (b) and (d), where one can observe differences between the lengths of the separation bubbles, especially on the suction side. For the cooled wall, a smaller recirculation region is noticed.
\begin{figure}[H]
	\centering
	\subfigure[Skin friction coefficient $c_f$]{\includegraphics[width=.49\linewidth]{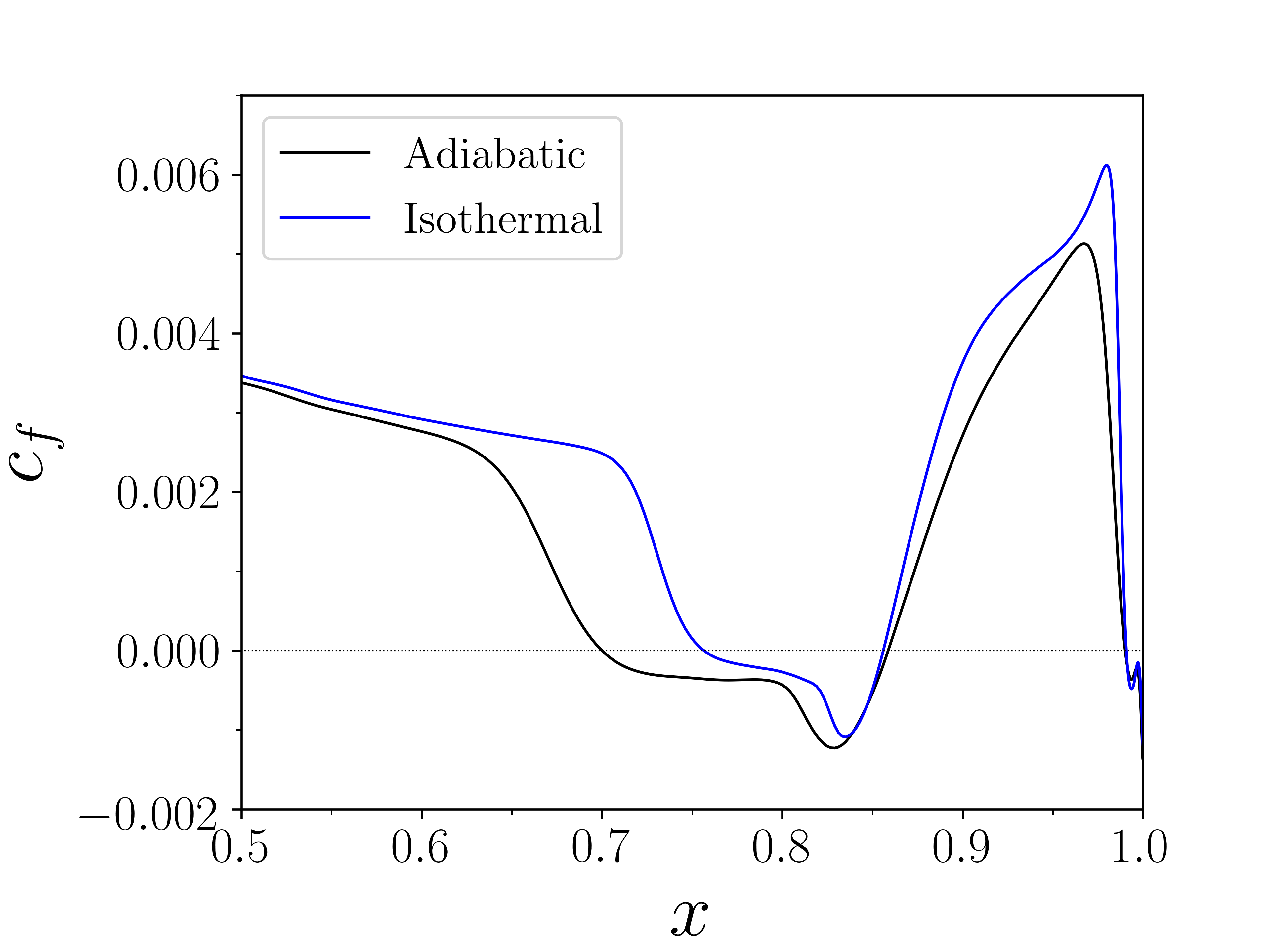}}
	\subfigure[Pressure coefficient $c_p$.]{\includegraphics[width=.49\linewidth]{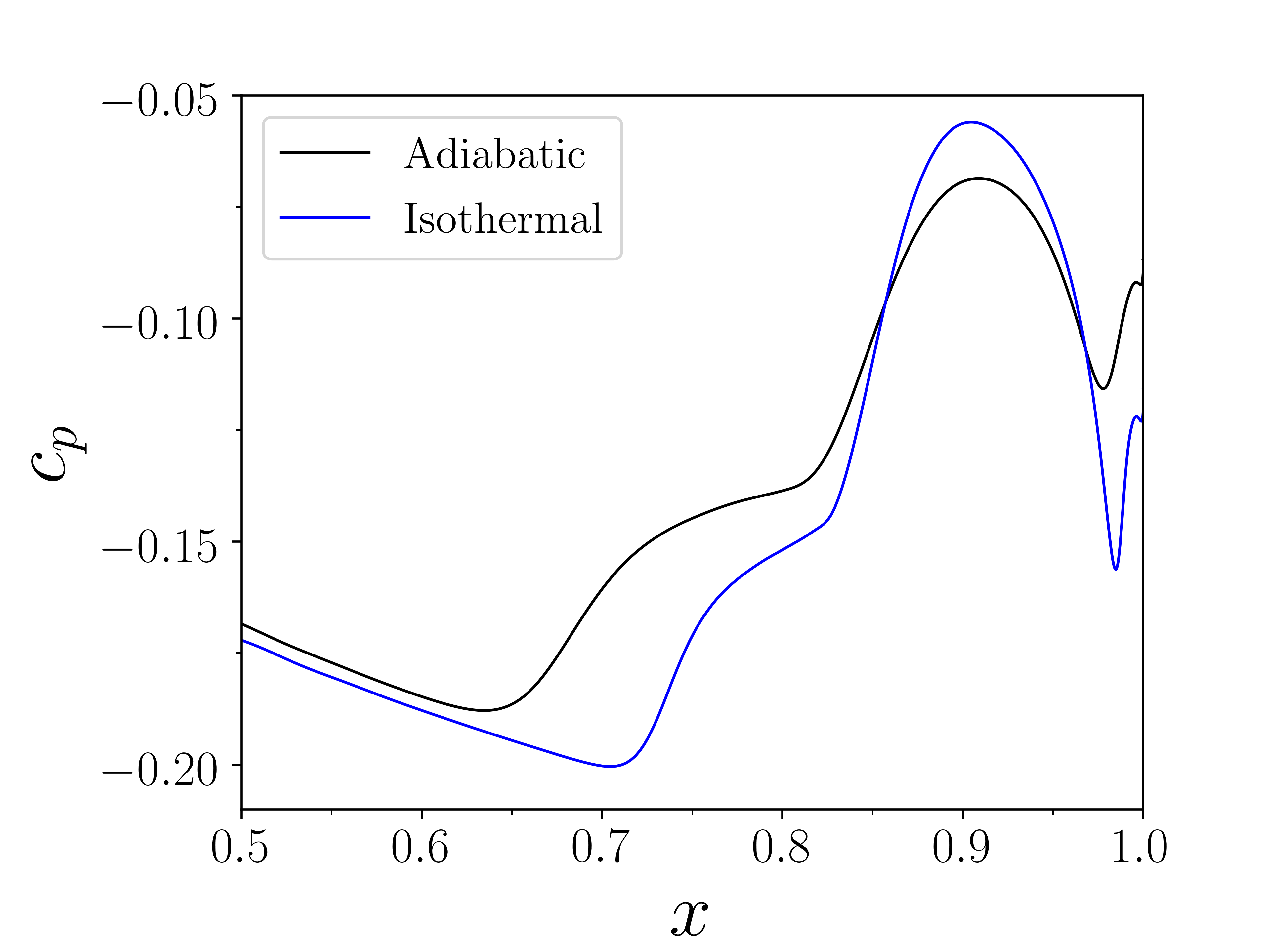}}
	\caption{Mean skin-friction and pressure coefficient distributions for the adiabatic (black) and cooled (blue) wall cases. The distributions are shown only along the suction side.}
	\label{fig:coefficients}
\end{figure}

\begin{figure}[H]
	\centering
	\begin{overpic}[trim = 40mm 2mm 40mm 2mm, clip,width=.49\linewidth]{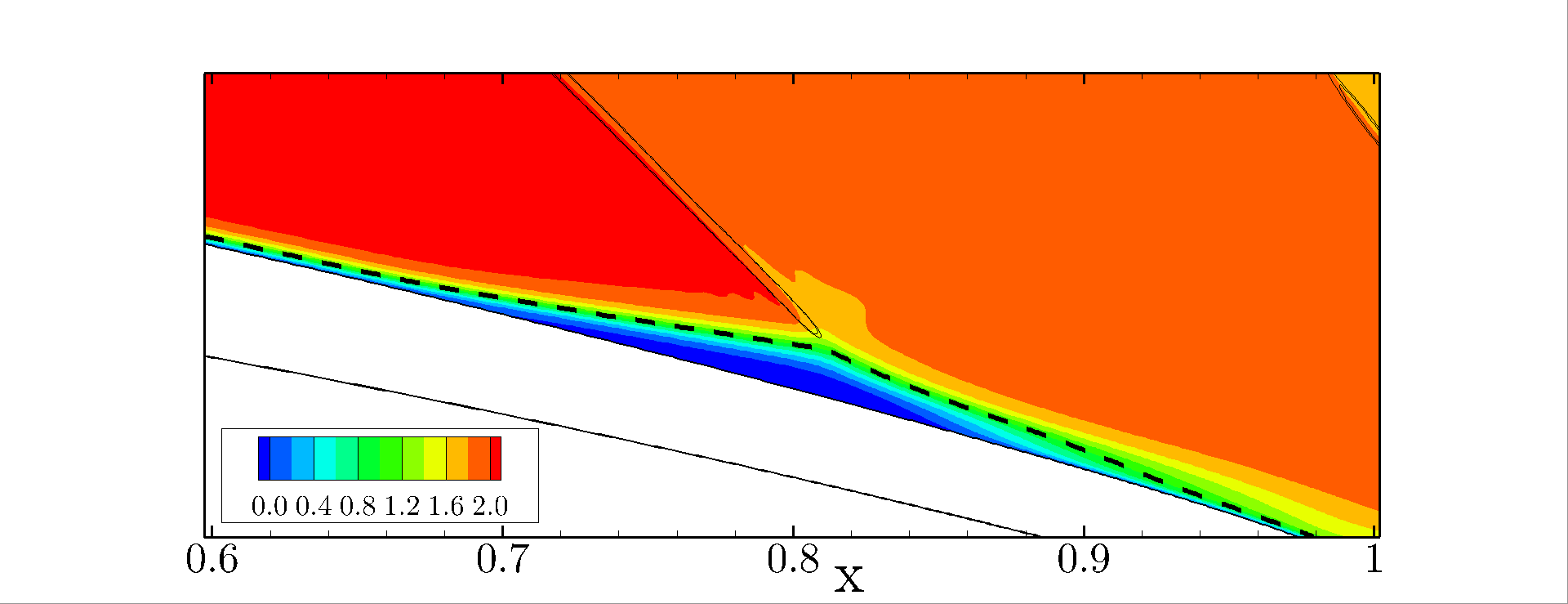}
		\put(0,34){(a)}
	\end{overpic}
	\begin{overpic}[trim = 40mm 2mm 40mm 2mm, clip,width=.49\linewidth]{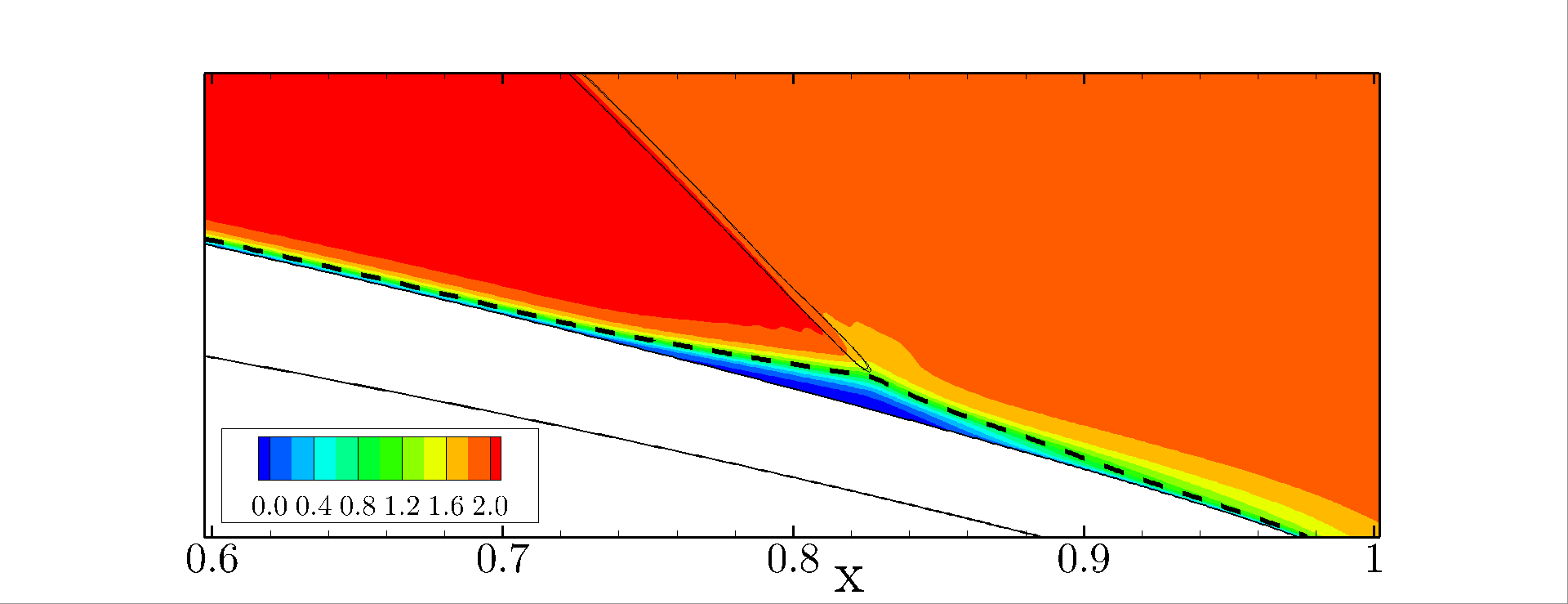}
		\put(0,34){(b)}
	\end{overpic}
	\begin{overpic}[trim = 40mm 2mm 40mm 2mm, clip,width=.49\linewidth]{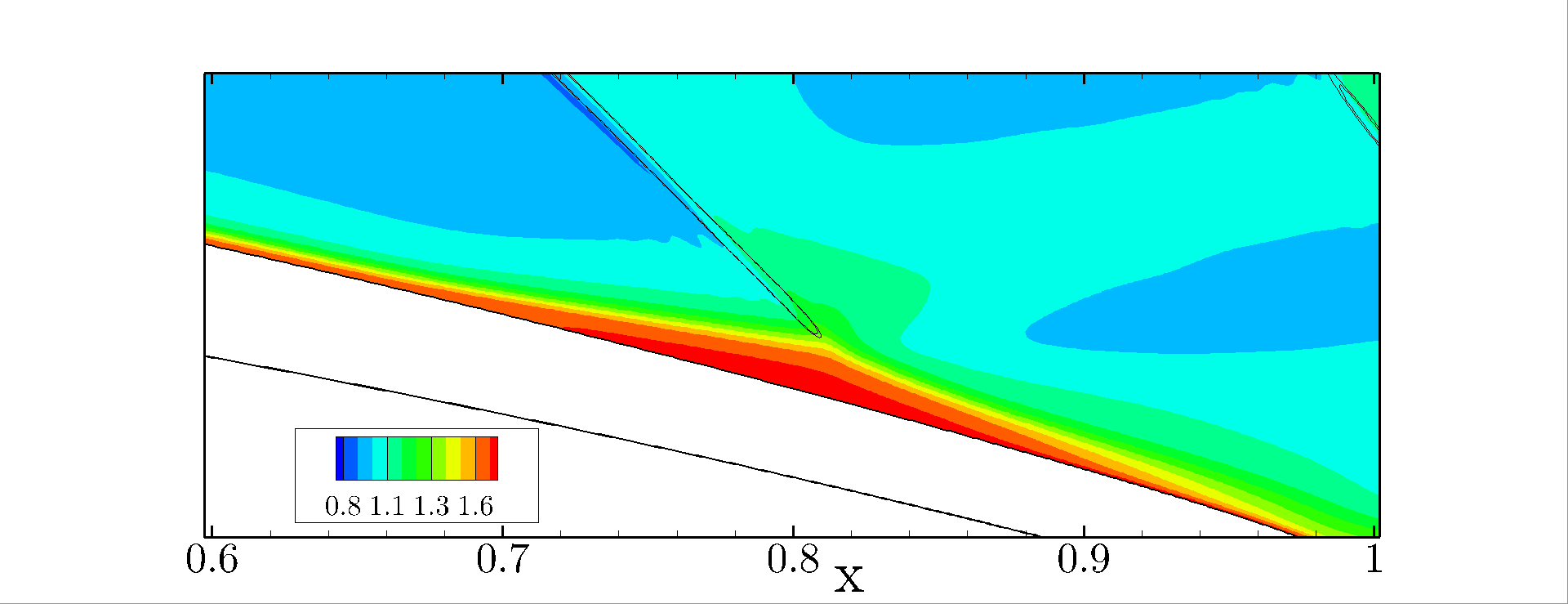}
		\put(0,34){(c)}
	\end{overpic}
	\begin{overpic}[trim = 40mm 2mm 40mm 2mm, clip,width=.49\linewidth]{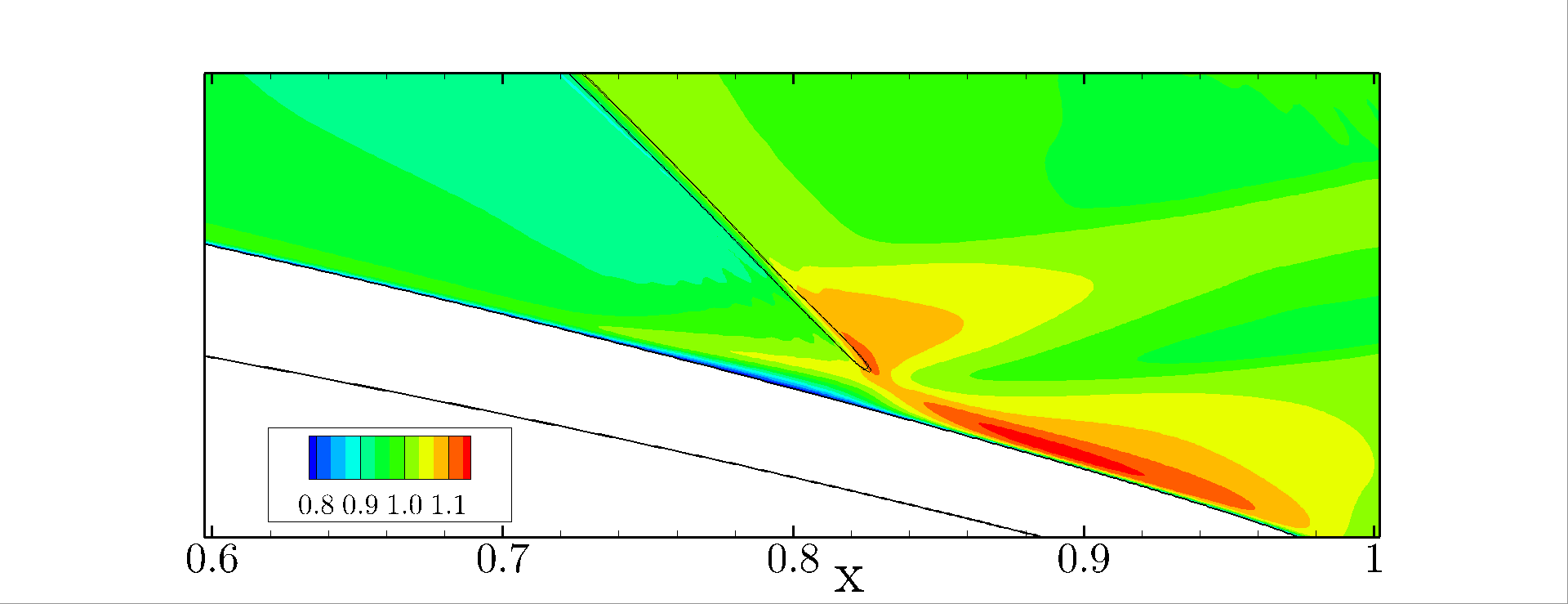}
		\put(0,34){(d)}
	\end{overpic}
	\caption{Time-averaged contours of normalized $u$-velocity (top) and temperature (bottom) for the adiabatic (left) and cooled (right) wall cases. The black lines display the shock waves visualized by pressure gradient magnitude. The black dashed lines show the sonic line.}
	\label{fig:mean_flows}
\end{figure}

The mean skin-friction coefficient distribution $c_f = \frac{\tau_w}{0.5 \rho_\infty U_{\infty}^2}$ is provided in Fig. \ref{fig:coefficients}(a) for the blade suction side. This plot shows the presence of a separation bubble characterized by locations where $c_f < 0$, which is delimited by a horizontal dashed line. The effect of cooling on the size of the recirculation region is evident. For the isothermal case, one can observe a downstream displacement of the separation region compared to the adiabatic wall setup. On the other hand, the reattachment locations are similar for both cases. Hence, the cooled wall depicts a smaller separation bubble. For the adiabatic wall case, the time-averaged  characteristic length of the separation bubble is $\langle L_{SB} \rangle = 0.16c_{ax}$ and it is observed along $0.70 < x/c_{ax} < 0.86$. For the cooled wall, $\langle L_{SB} \rangle = 0.10c_{ax}$ and it is formed on $0.75 < x/c_{ax} < 0.85$. After a small negative skin-friction coefficient plateau, a similar recovery is observed downstream of the reattachment location for both cases.

Figure \ref{fig:coefficients} (b) plots the mean pressure coefficient $c_p = \frac{p - p_\infty}{0.5 \rho_\infty U_{\infty}^2}$ along the airfoil chord. For both adiabatic and cold wall cases, it is possible to note two pressure rises: the first occurs near the separation point due to the compression waves formed upstream of the separation bubble, and the second takes place near the reattachment location as a result of the incident shock impingement and the turbulence amplification mechanism \cite{fang2020}. For the cooled wall setup, a steeper variation of $c_p$ is observed, especially for the second pressure rise.

To highlight the influence of the wall thermal boundary conditions on the size and shape of the separation bubbles, the mean (spanwise and time averaged) $u$-velocity contours are presented in Figs. \ref{fig:mean_flows} (a) and (b), for the adiabatic and isothermal cases, respectively. Here, the velocity component is normalized by the inlet speed of sound. These figures reinforce the findings observed in the friction coefficient distributions. The main effect of wall cooling is to reduce the viscous length scales near the wall \cite{Bernardini2016,Volpiani2018} which in turn affects the shock penetration, as shown in \ref{fig:mean_flows} (a) and (b). One can see that the impinging shock penetrates deeper in the boundary layer for the cooled wall case due to the displacement of the sonic line (displayed as a dashed line) towards the wall. This effect is responsible for the steeper variation in the pressure coefficient observed in Fig. \ref{fig:coefficients}(b). One can also see that, for the cooled wall, the incident shock reaches further downstream compared to the adiabatic case.

Figures \ref{fig:mean_flows} (c) and (d) show the mean temperature fields for the the adiabatic and cold wall boundary conditions, respectively. The values are presented normalized by the inlet temperature. For the former case, one can observe that a region of maximum temperature occurs within the separation bubble. On the other hand, when cooling is applied, higher temperature values are observed in the free shear layer, downstream the bubble. For the adiabatic wall, friction from the shear stresses near the wall and around the bubble are converted into heat which is transferred along the boundary layer and inside the bubble. This causes the near-wall flow to reach higher temperatures. 
%On the other hand, heat from the flow is transferred to the blade in the isothermal case, which has a lower temperature than the surrounding flow.
However, heat from the flow is transferred to the blade in the isothermal case, which has a lower temperature than the surrounding flow. For the cooled wall case, the maximum temperature values are observed along the free shear layer, behind the bubble, due to strong shearing effects that cause aerodynamic heating. 

%\red{A mudança de pressão pós-bolha não pode ter relação com os diferentes campos de temperatura tbm? $P = \rho R T$...}
%
\begin{figure}[H]
	\begin{center}
		{\includegraphics[trim = 1mm 1mm 1mm 1mm, clip, width=0.97\textwidth]{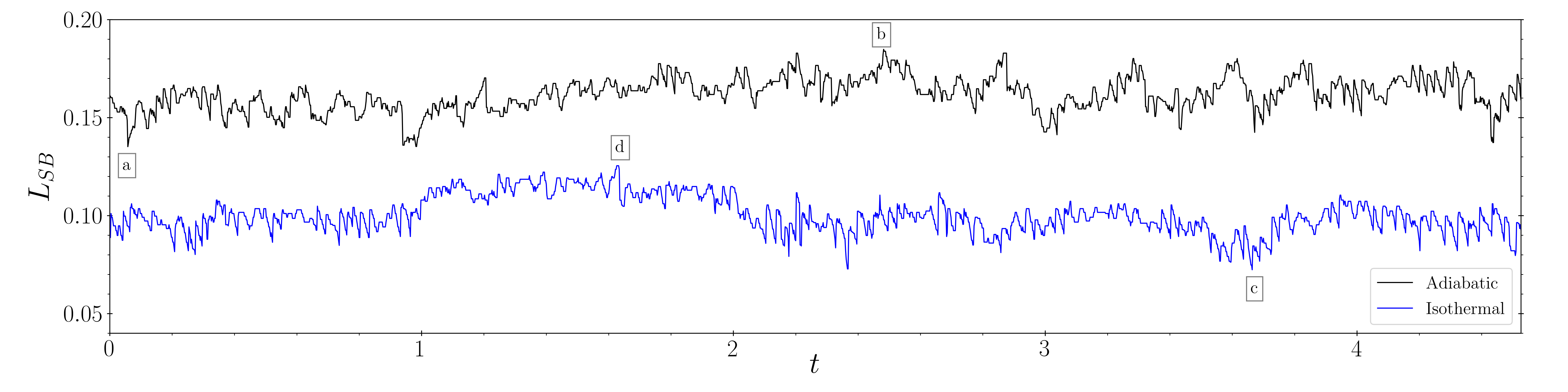}}
	\end{center}
	\caption{Temporal variation of the suction side separation bubble length $L_{SB}$ for the adiabatic (top) and cooled (bottom) wall cases.}
	\label{fig:LSB}
\end{figure}
\begin{figure}
	\centering
	\begin{overpic}[trim = 40mm 20mm 40mm 50mm,clip,width=0.48\textwidth]{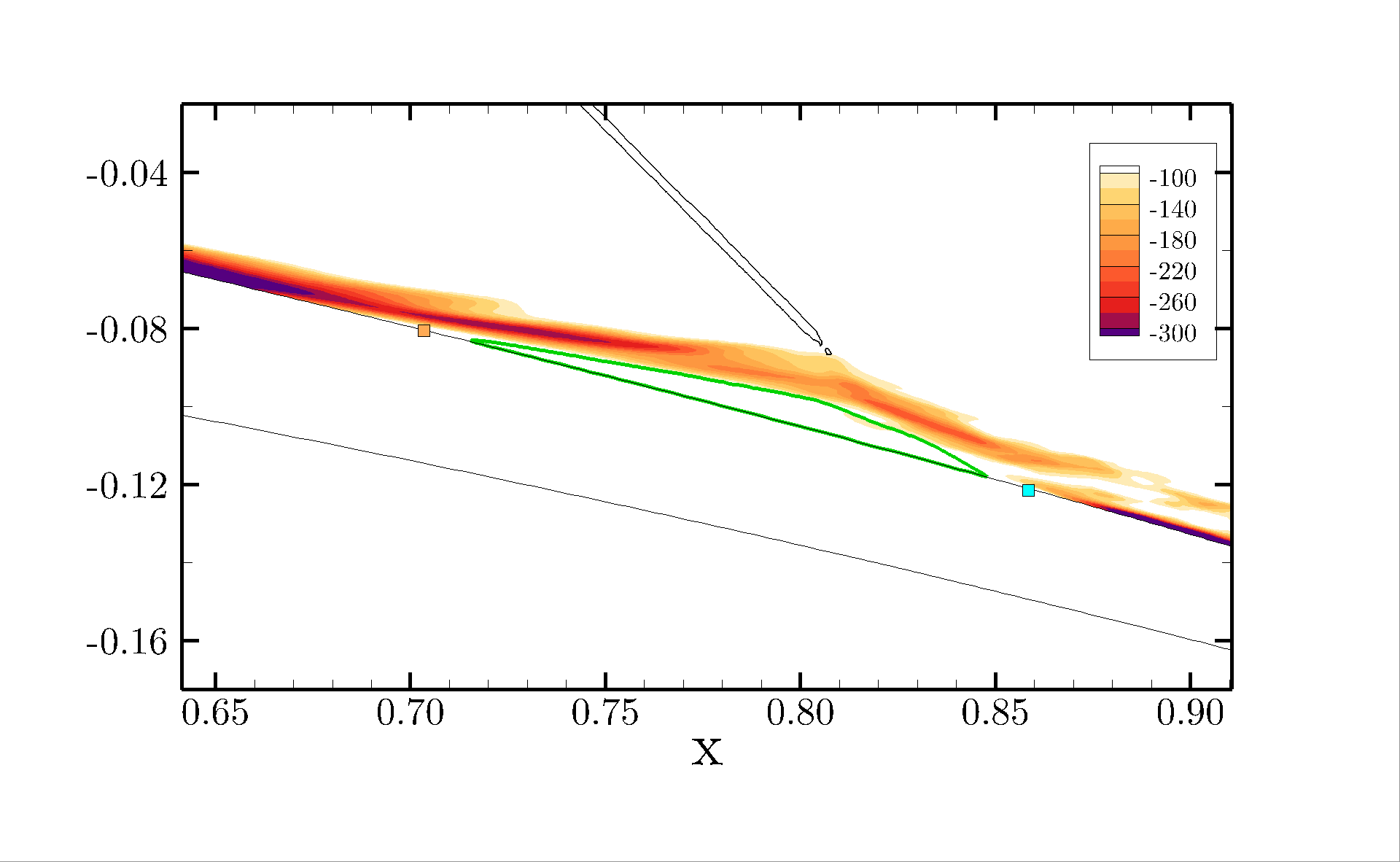}
		\put(10,52){(a)}
	\end{overpic} 
	\begin{overpic}[trim = 40mm 20mm 40mmm 50mm,clip,width=0.48\textwidth]{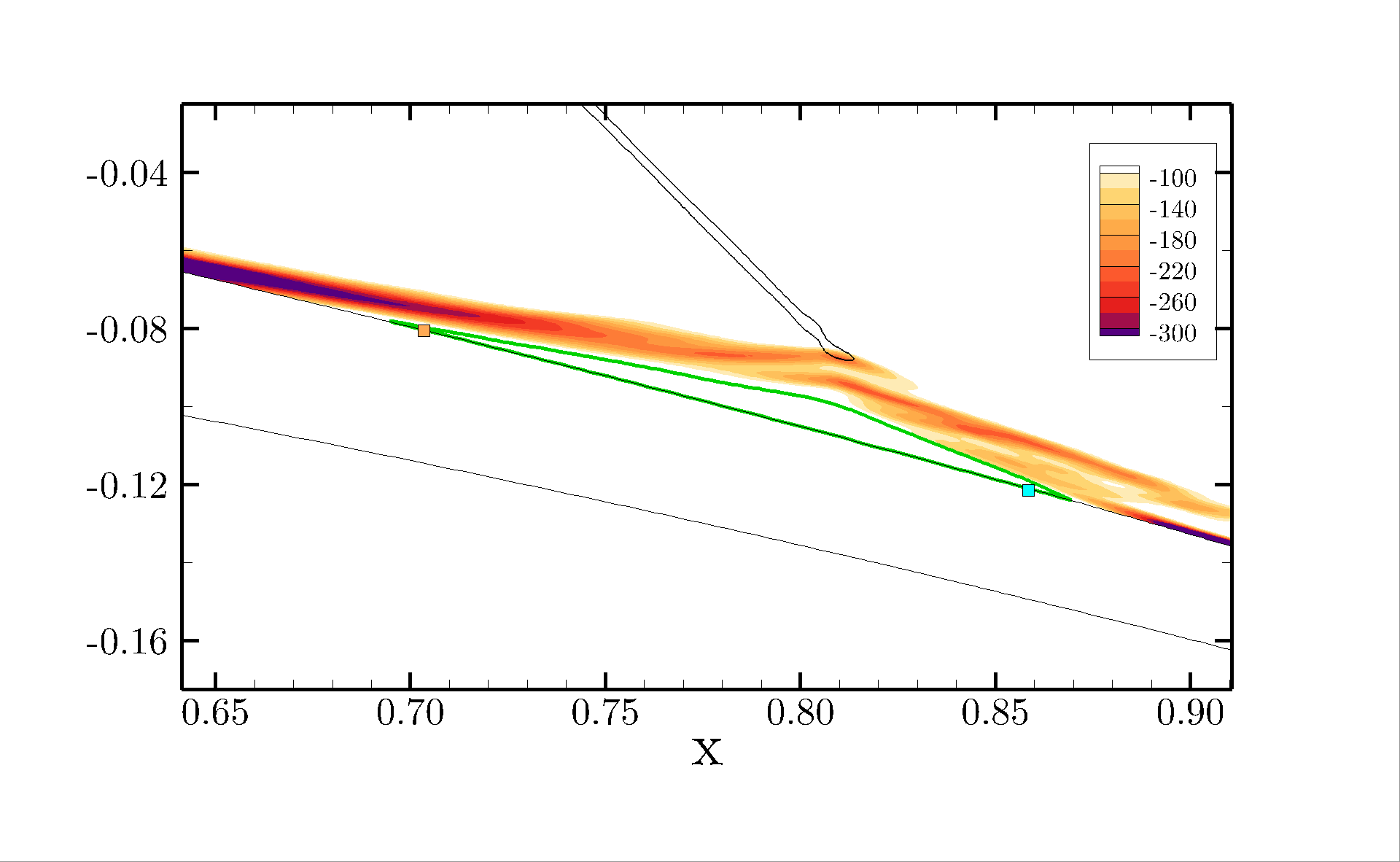}
		\put(10,52){(b)}
	\end{overpic} 
	\begin{overpic}[trim = 40mm 20mm 40mm 50mm,clip,width=0.48\textwidth]{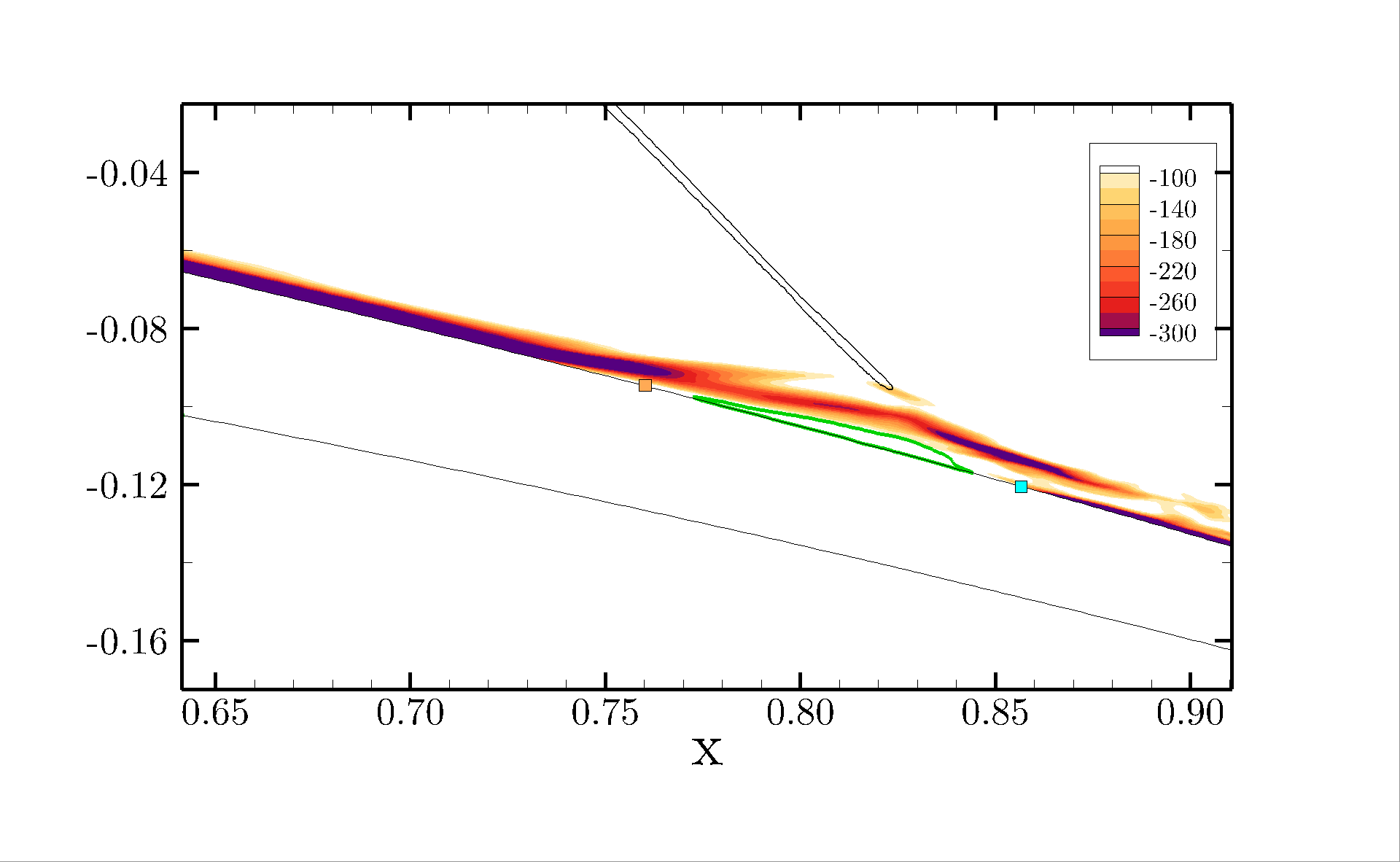}
		\put(10,52){(c)}
	\end{overpic} 
	\begin{overpic}[trim = 40mm 20mm 40mm 50mm,clip,width=0.48\textwidth]{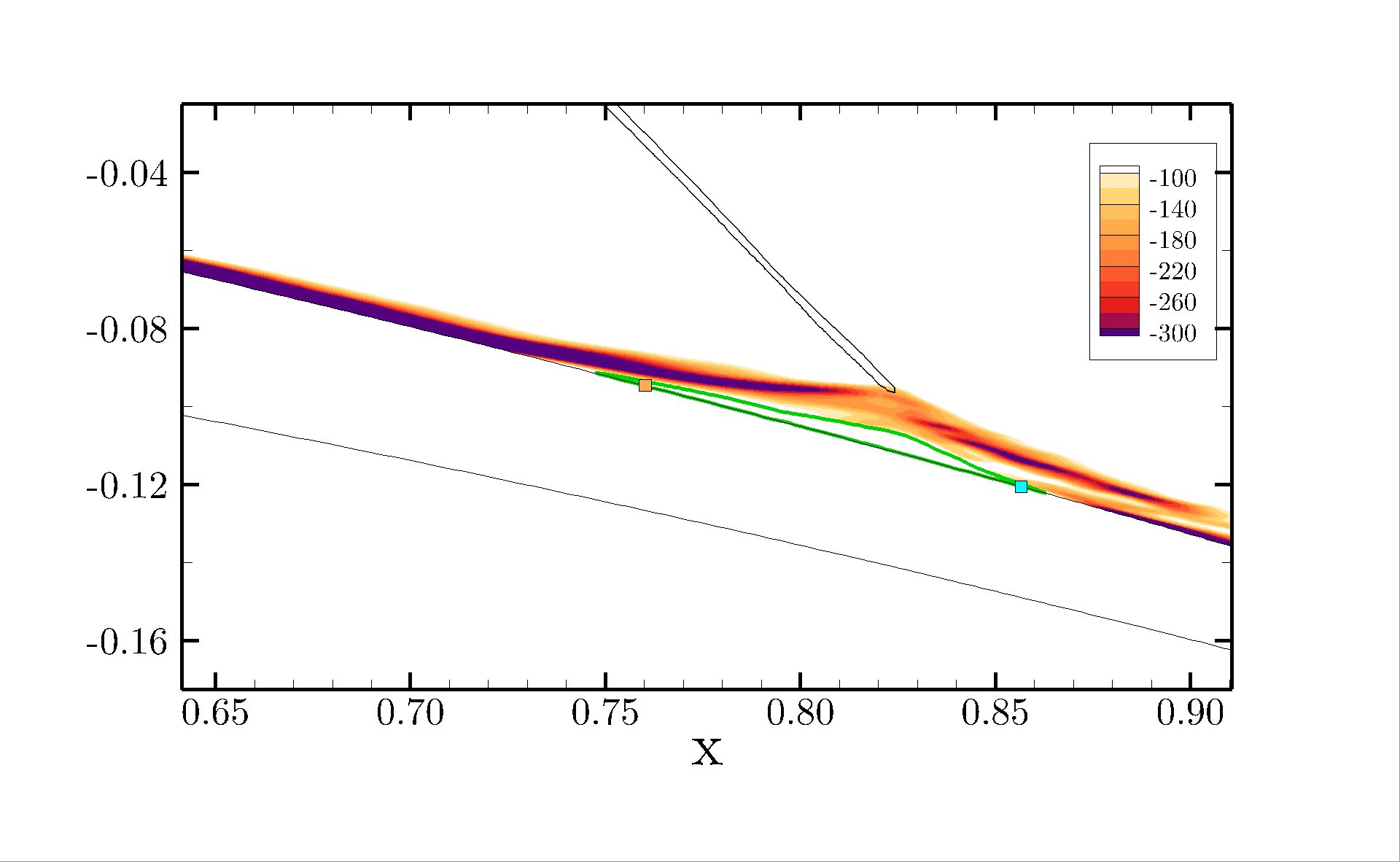}
		\put(10,52){(d)}
	\end{overpic} 
	\caption{Spanwise $z$-vorticity contours at different time instants for the adiabatic (top) and cooled (bottom) wall cases. The green line delimits the bubble while the black line shows the incident shock wave.}
	\label{fig:z_vorticity}
\end{figure}

The temporal evolution of the separation bubble length $L_{SB}$ is shown in Fig. \ref{fig:LSB} for the adiabatic and isothermal walls. The instantaneous length of the bubble is defined as the distance between the instantaneous reattachment and separation locations. One can observe that the separation region undergoes a contraction/expansion motion for both cases. The excursions from the mean appear to be similar for both cases. A spectral analysis of this signal should provide further information on the frequency scales related to the bubble motion and such analysis should be conducted in future work, once longer signals are collected for statistical convergence of the lower frequencies of interest.

To highlight the 2D structure of the suction side separation bubble and shear layer at different time instants, snapshots of $z$-vorticity are displayed in Fig. \ref{fig:z_vorticity} for both thermal boundary conditions. These snapshots correspond to the instants indicated by the letters ``a-d'' in Fig. \ref{fig:LSB}. The region enclosed by the green line shows the separation region and the black lines display the impinging shocks. In addition, the mean separation and reattachment positions are indicated by the orange and cyan squares, respectively. For both cases, when the bubble suffers a contraction, the instantaneous separation (reattachment) point moves downstream (upstream) with respect to its mean value, as can be visualized in Figs. \ref{fig:z_vorticity}(a) and (c). On the other hand, when the bubble undergoes an expansion, one can observe the upstream (downstream) movement of the instantaneous separation (reattachment) point with respect to its mean position. This indicates that the bubble has a breathing pattern, but its central position does not have large excursions from the mean. Figure \ref{fig:z_vorticity} also shows that the shear layer downstream of the bubble is more diffused for the adiabatic case, while more concentrated vorticity values are observed when cooling is applied. These findings corroborate the maximum temperature values observed in Figs. \ref{fig:mean_flows}(c) and (d). For example, in the adiabatic case, the shear layer around the bubble creates a zone of intense heating.
\begin{figure}
	\centering
	\begin{overpic}[trim = 40mm 2mm 40mm 2mm, clip,width=.33\linewidth]{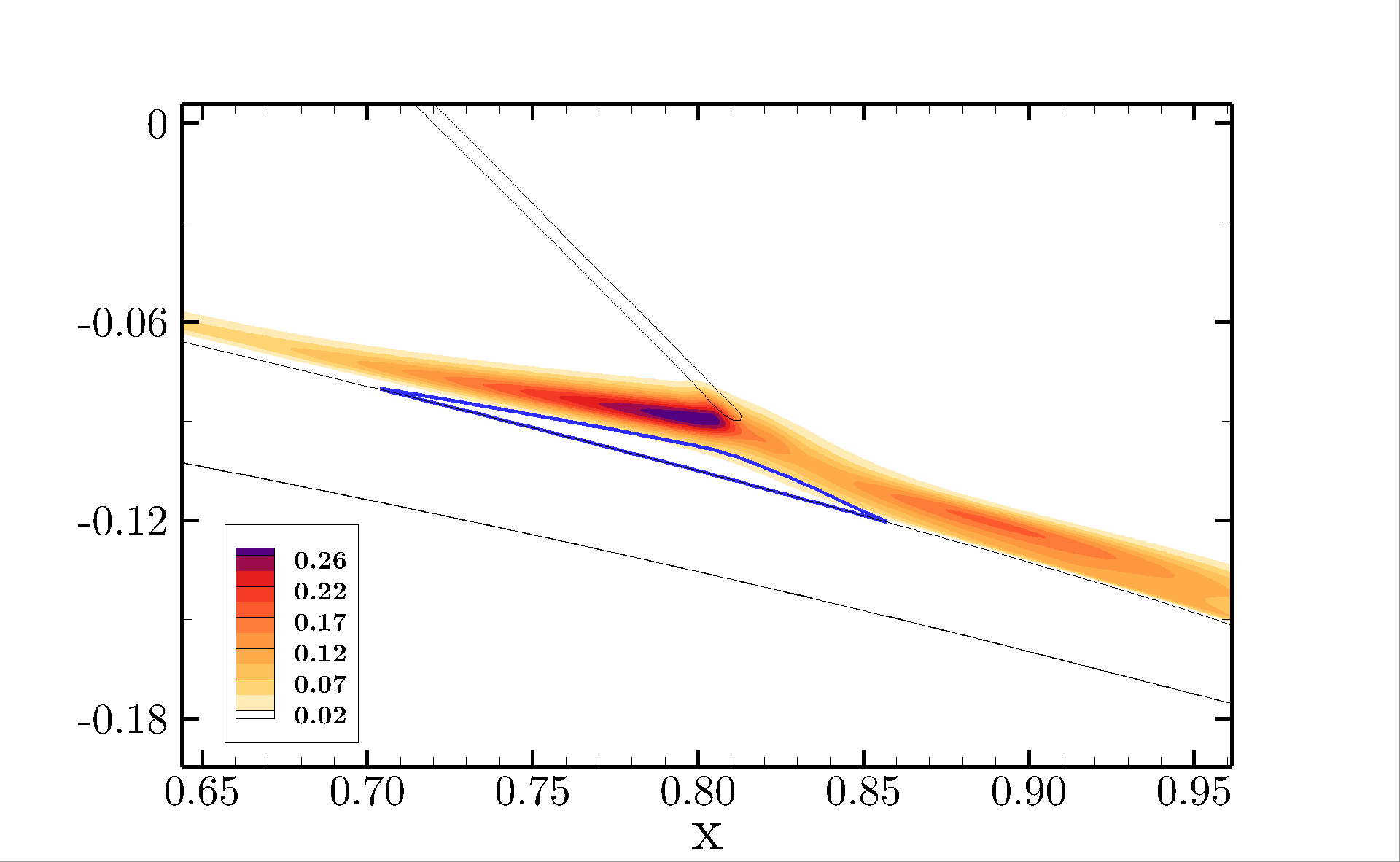}
		\put(10,54){(a)}
	\end{overpic}
	\begin{overpic}[trim = 40mm 2mm 40mm 2mm, clip,width=.33\linewidth]{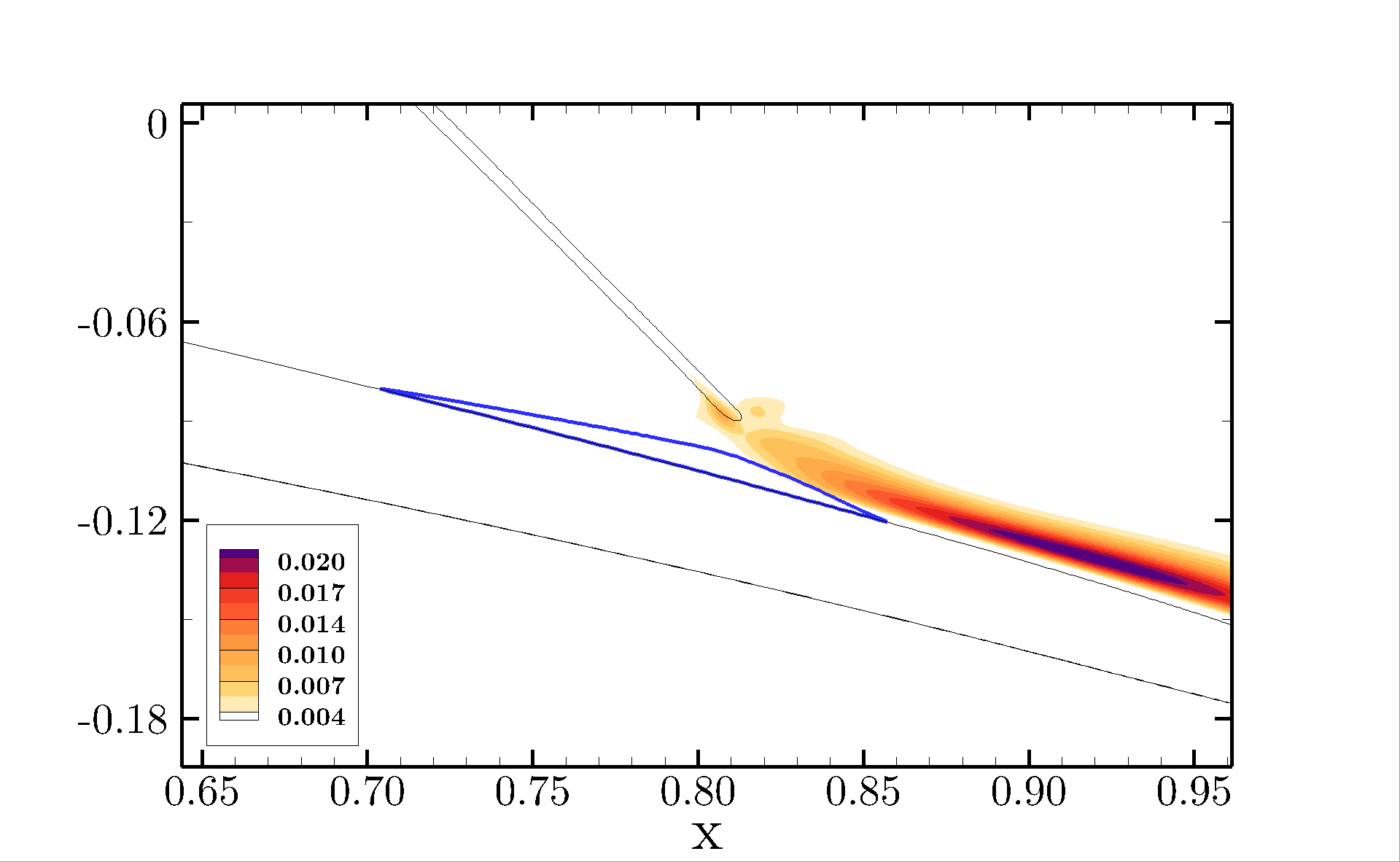}
		\put(10,54){(b)}
	\end{overpic}
	\begin{overpic}[trim = 40mm 2mm 40mm 2mm, clip,width=.33\linewidth]{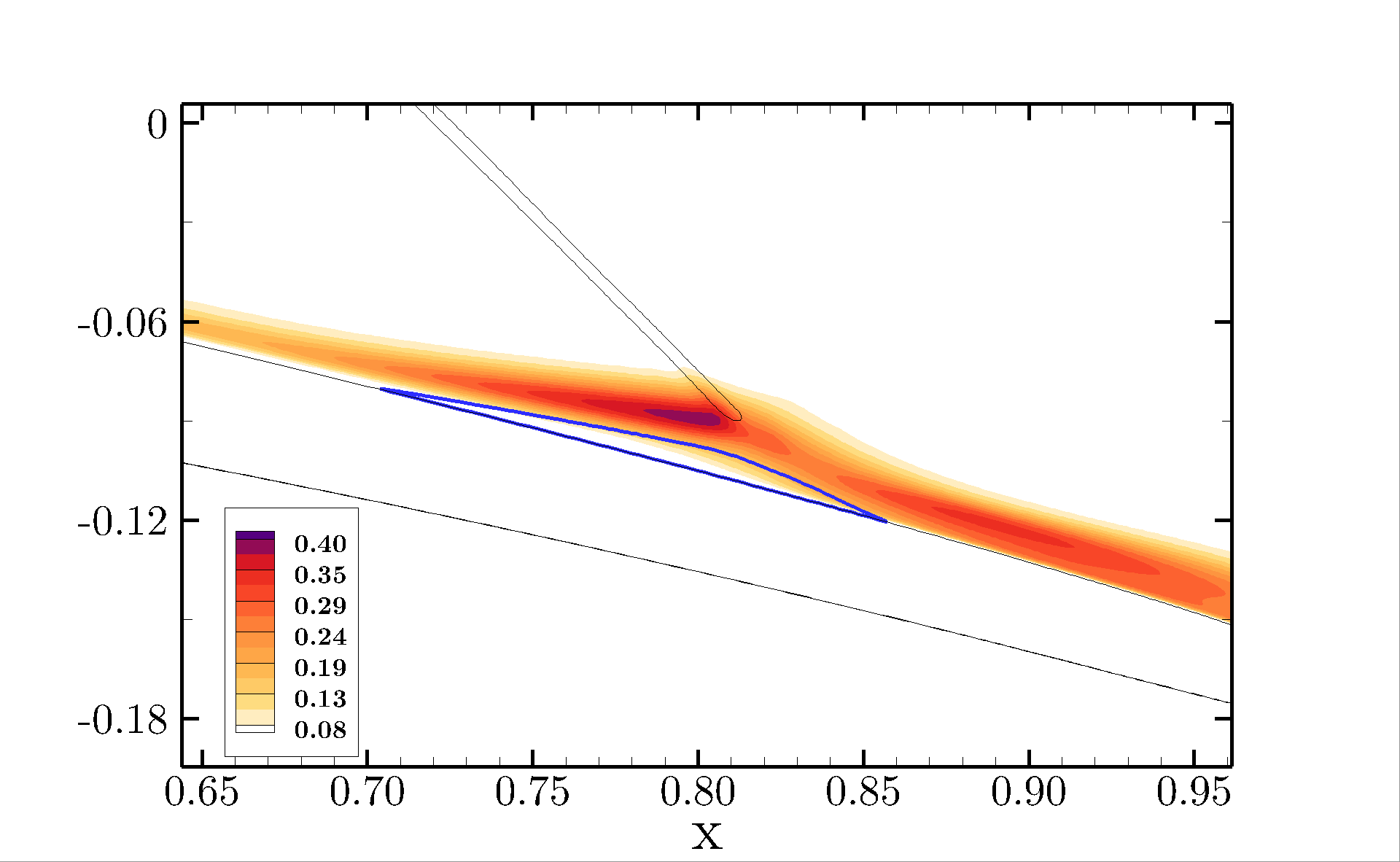}
		\put(10,54){(c)}
	\end{overpic}
	\begin{overpic}[trim = 40mm 2mm 40mm 2mm, clip,width=.33\linewidth]{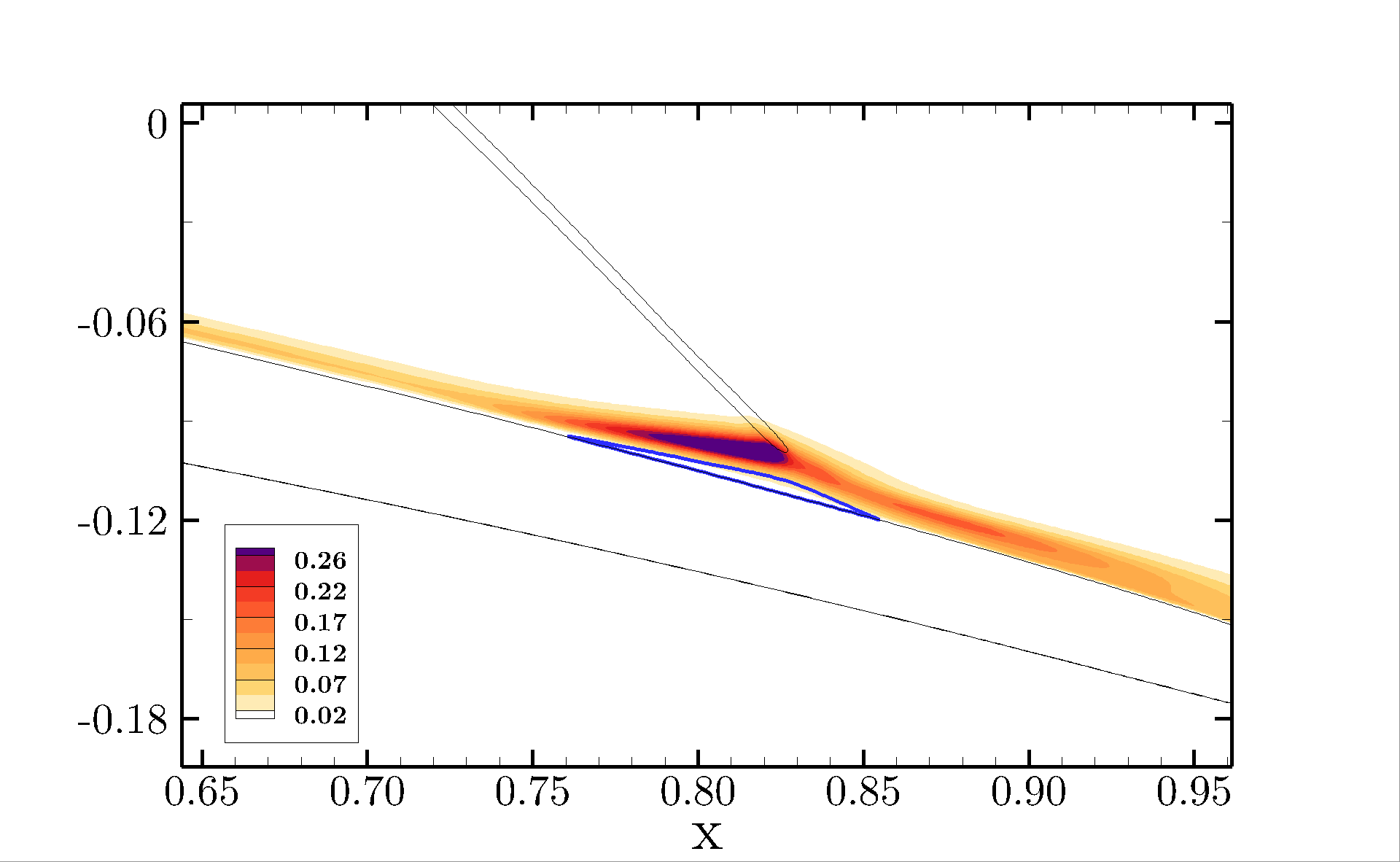}
		\put(10,54){(d)}
	\end{overpic}
	\begin{overpic}[trim = 40mm 2mm 40mm 2mm, clip,width=.33\linewidth]{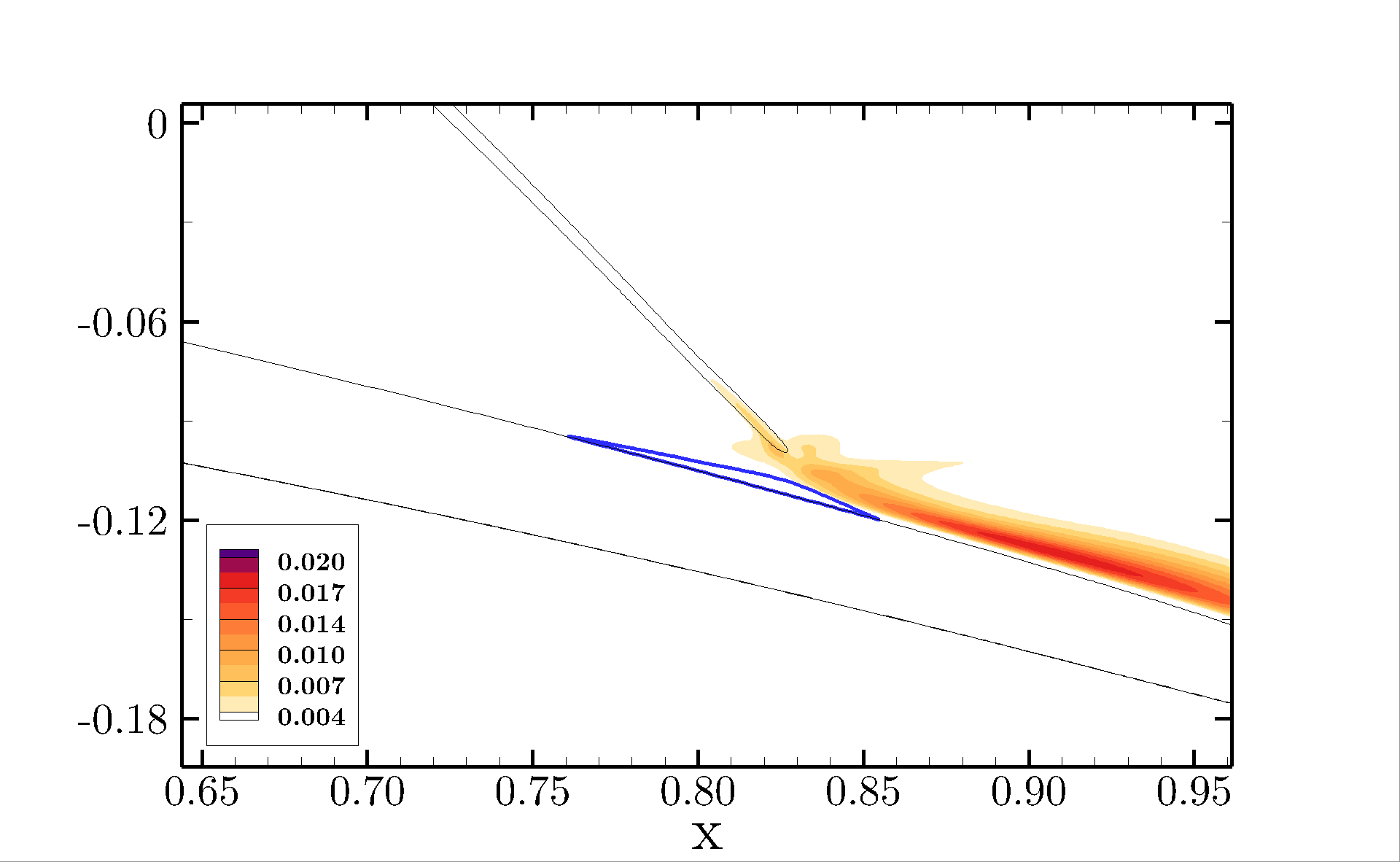}
		\put(10,54){(e)}
	\end{overpic}
	\begin{overpic}[trim = 40mm 2mm 40mm 2mm, clip,width=.33\linewidth]{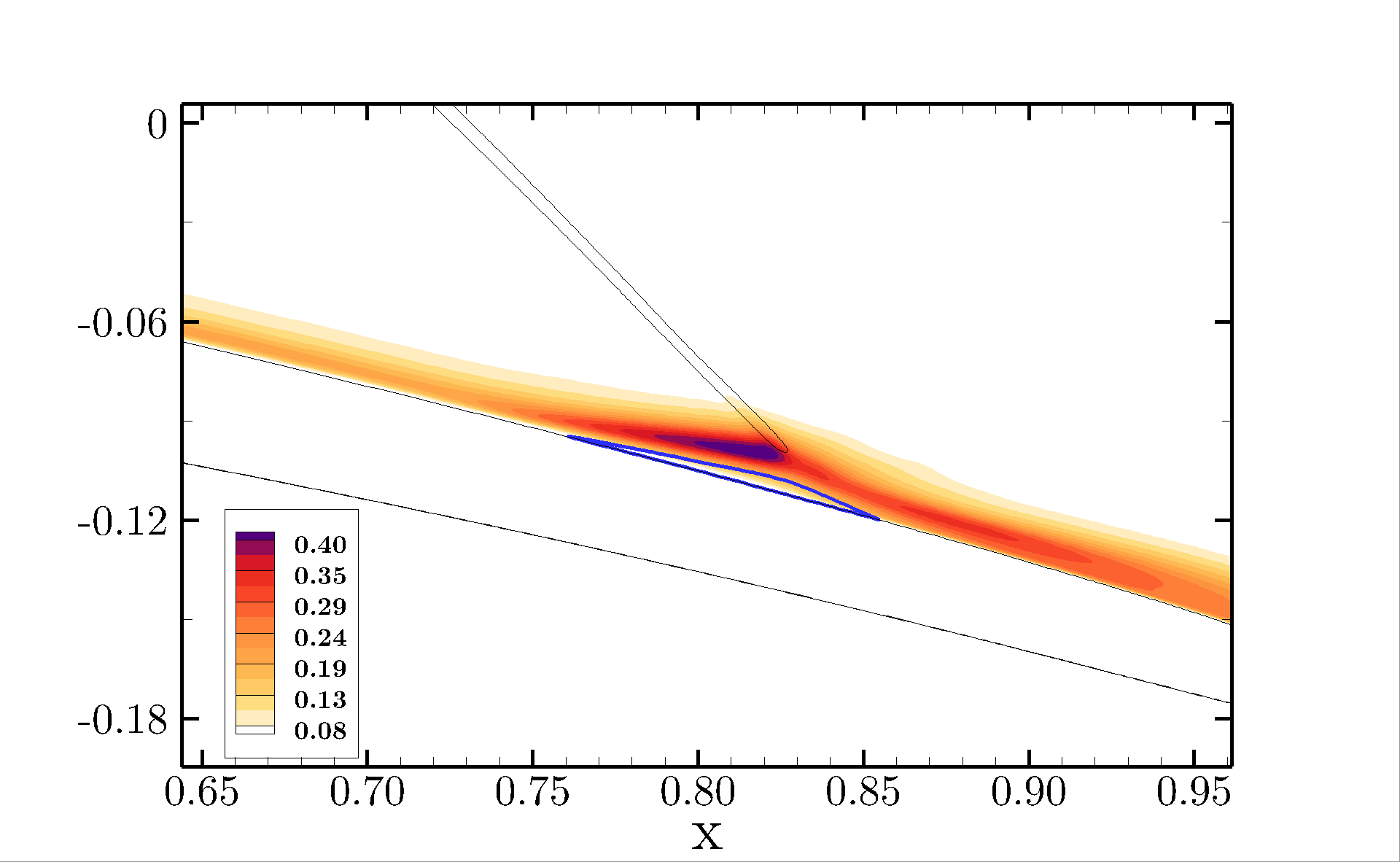}
		\put(10,54){(f)}
	\end{overpic}
	\caption{Turbulence quantities on the suction side for the adiabatic (top) and cooled (bottom) cases: $\langle u_t u_t \rangle$ (left), $\langle u_n u_n \rangle$ (middle) and TKE (right). }
	\label{fig:turbulence_quantities}
\end{figure}

The effects of the thermal boundary conditions on the turbulence properties, are investigated by the tangential and wall-normal Reynolds stresses, $\langle u_t u_t \rangle$ and $\langle u_n u_n \rangle$, respectively, and the turbulent kinetic energy (TKE) are presented in Fig. \ref{fig:turbulence_quantities}. In this figure, the top and bottom rows display results for the adiabatic and isothermal walls, respectively. In Figs. \ref{fig:turbulence_quantities} (a) and (d), it can be seen that the highest fluctuations of $\langle u_t u_t \rangle$ are observed just upstream of the shock-bubble interaction for both cases, with similar fluctuation values. The amplification of $\langle u_t u_t \rangle$ is associated with the development of the shear layer \cite{fang2020}. The peak values of $\langle u_n u_n \rangle$ are found along the free shear layer downstream of the bubble. The magnitude of $\langle u_n u_n \rangle$ decreases when cooling is applied. In Figs. \ref{fig:turbulence_quantities} (c) and (f), one can observe that the turbulent kinetic energy combines the trends observed from the $\langle u_t u_t \rangle$ and $\langle u_n u_n \rangle$ components. In addition, before the SBLI, we can notice a downstream displacement of the maximum turbulence amplification location for the cooled wall case. This occurs due to the higher shock penetration discussed previously.

\section{Conclusions}

Wall-resolved large eddy simulations are employed to investigate thermal effects in a supersonic turbine cascade. Simulations are performed for adiabatic and isothermal boundary conditions, where in the latter case the blade is cooled. For the present flow configurations, oblique shock waves are generated at the leading edges of the airfoils, and they interact with the boundary layers of the neighboring blades. A study of the shock-boundary layer interactions is presented for the blade suction side, where an incident oblique shock reflects on the wall leading to the formation of a separation bubble.

The impact of the thermal boundary conditions on the separation bubbles is investigated. The distributions of mean skin-friction show that the separation bubble is considerably smaller for the cooled wall compared to the adiabatic case. Pressure coefficient distributions show that a steeper pressure rise occurs downstream the incident shock wave for the cooled wall. For this case, cooling induces the formation of a thinner boundary layer and the sonic line forms closer to the wall. Results in terms of mean velocity contours reveal that the more pronounced pressure rise occurs due to the higher penetration of the incident shock in the isothermal (cooled) wall. Maximum temperature values are observed along the bubble, for the adiabatic case, and the free shear layer, for the cooled wall. In the former case, aerodynamic heating is transferred to the bubble due to a surrounding shear layer. For the latter case, intense shearing is observed along the free shear layer, behind the bubble, and leads to high temperatures. 

An analysis of the instantaneous separation and reattachment locations demonstrates that the separation bubbles have a breathing pattern of contractions and expansions. For the contraction motions, the instantaneous separation point moves downstream while the reattachment point moves upstream. The other way around is observed for the expansion motions. The tangential Reynolds stress distributions reach maximum values just upstream the shock-bubble interactions, being similar for both the adiabatic and isothermal walls. However, due to the higher shock penetration of the isothermal wall, the peaks appear more downstream along the blade chord. The wall-normal Reynolds stresses reach maximum amplitudes downstream the SBLI and they are more pronounced for the adiabatic wall. In future work, further analysis of the SBLI dynamics will be provided for both suction and pressure side boundary layers.

\section*{Acknowledgments}
The authors acknowledge the financial support received from Funda\c{c}\~ao de Amparo \`a Pesquisa do Estado de S\~ao Paulo, FAPESP, under grants No.\ 2013/08293-7, 2019/26196-5 and 2021/06448-0. The authors also thank Conselho Nacional de Desenvolvimento Científico e Tecnológico, CNPq, for supporting this research under grants No.\ 407842/2018-7 and 308017/2021-8. This work was granted access to the HPC resources of IDRIS under the allocation 2021-A0112A12067 made by GENCI.

\bibliography{sample}

\begin{thebibliography}{21}
\newcommand{\enquote}[1]{``#1''}
\providecommand{\natexlab}[1]{#1}
\providecommand{\url}[1]{\texttt{#1}}
\providecommand{\urlprefix}{URL }
\expandafter\ifx\csname urlstyle\endcsname\relax
  \providecommand{\doi}[1]{\discretionary{}{}{}https://doi.org/#1}\else
  \providecommand{\doi}[1]{\discretionary{}{}{}\urlstyle{rm}\url{https://doi.org/#1}}\fi

\bibitem[{Paniagua et~al.(2014)Paniagua, Iorio, Vinha, and
  Sousa}]{PANIAGUA201465}
Paniagua, G., Iorio, M., Vinha, N., and Sousa, J., \enquote{Design and analysis
  of pioneering high supersonic axial turbines,} \emph{International Journal of
  Mechanical Sciences}, Vol.~89, 2014, pp. 65 -- 77.

\bibitem[{Babinsky and Harvey(2011)}]{babinsky_harvey_2011}
Babinsky, H., and Harvey, J., \emph{Shock Wave-Boundary-Layer Interactions},
  Cambridge Aerospace Series, Cambridge University Press, 2011.

\bibitem[{Gaitonde(2015)}]{GAITONDE2015}
Gaitonde, D.~V., \enquote{Progress in shock wave/boundary layer interactions,}
  \emph{Progress in Aerospace Sciences}, Vol.~72, 2015, pp. 80--99.

\bibitem[{Lui et~al.(2021)Lui, Wolf, Braun, Rahbari, and Paniagua}]{Lui2021}
Lui, H., Wolf, W.~R., Braun, J., Rahbari, I., and Paniagua, G.,
  \enquote{Numerical simulations of supersonic stator cascades: Assessment of
  {LES} and {RANS} calculations,} \emph{AIAA Aviation Forum}, 2021.

\bibitem[{Lui et~al.(2022)Lui, Ricciardi, Wolf, Braun, Rahbari, and
  Paniagua}]{lui2022}
Lui, H. F.~S., Ricciardi, T.~R., Wolf, W.~R., Braun, J., Rahbari, I., and
  Paniagua, G., \enquote{Shock boundary layer interactions in supersonic
  turbine cascades,} \emph{ArXiv e-prints, arXiv:2204.08065 [physics.flu-dyn]},
  2022.

\bibitem[{Schülein(2006)}]{Erich2006}
Schülein, E., \enquote{Skin Friction and Heat Flux Measurements in
  Shock/Boundary Layer Interaction Flows,} \emph{AIAA Journal}, Vol.~44, No.~8,
  2006, pp. 1732--1741.

\bibitem[{Jaunet et~al.(2014)Jaunet, Debiève, and Dupont}]{Jaunet2014}
Jaunet, V., Debiève, J.~F., and Dupont, P., \enquote{Length Scales and Time
  Scales of a Heated Shock-Wave/Boundary-Layer Interaction,} \emph{AIAA
  Journal}, Vol.~52, No.~11, 2014, pp. 2524--2532.

\bibitem[{Bernardini et~al.(2016)Bernardini, Asproulias, Larsson, Pirozzoli,
  and Grasso}]{Bernardini2016}
Bernardini, M., Asproulias, I., Larsson, J., Pirozzoli, S., and Grasso, F.,
  \enquote{Heat transfer and wall temperature effects in shock wave turbulent
  boundary layer interactions,} \emph{Phys. Rev. Fluids}, Vol.~1, 2016, p.
  084403.

\bibitem[{Volpiani et~al.(2018)Volpiani, Bernardini, and
  Larsson}]{Volpiani2018}
Volpiani, P.~S., Bernardini, M., and Larsson, J., \enquote{Effects of a
  nonadiabatic wall on supersonic shock/boundary-layer interactions,}
  \emph{Phys. Rev. Fluids}, Vol.~3, 2018, p. 083401.

\bibitem[{Nagarajan et~al.(2003)Nagarajan, Lele, and Ferziger}]{Nagarajan2003}
Nagarajan, S., Lele, S.~K., and Ferziger, J.~H., \enquote{A robust high-order
  compact method for large eddy simulation,} \emph{Journal of Computational
  Physics}, Vol. 191, No.~2, 2003, pp. 392--419.

\bibitem[{Beam and Warming(1978)}]{Beam1978}
Beam, R.~M., and Warming, R.~F., \enquote{An Implicit Factored Scheme for the
  Compressible Navier-Stokes Equations,} \emph{AIAA Journal}, Vol.~16, No.~4,
  1978, pp. 393--402.

\bibitem[{Delfs(2001)}]{Delfis2001}
Delfs, J., \enquote{An overlapped grid technique for high resolution {CAA}
  schemes for complex geometries,} \emph{7th AIAA/CEAS Aeroacoustics Conference
  and Exhibit}, 2001.

\bibitem[{Bhaskaran and Lele(2010)}]{Bhaskaran}
Bhaskaran, R., and Lele, S.~K., \enquote{Large eddy simulation of free-stream
  turbulence effects on heat transfer to a high-pressure turbine cascade,}
  \emph{Journal of Turbulence}, Vol.~11, 2010, p.~N6.

\bibitem[{Lele(1992)}]{Lele1992}
Lele, S.~K., \enquote{Compact finite difference schemes with spectral-like
  resolution,} \emph{Journal of Computational Physics}, Vol. 103, No.~1, 1992,
  pp. 16--42.

\bibitem[{Cook(2007)}]{Cook2007}
Cook, A.~W., \enquote{Artificial fluid properties for large-eddy simulation of
  compressible turbulent mixing,} \emph{Physics of Fluids}, Vol.~19, No.~5,
  2007, p. 055103.

\bibitem[{Kawai et~al.(2010)Kawai, Shankar, and Lele}]{Kawai2010}
Kawai, S., Shankar, S.~K., and Lele, S.~K., \enquote{Assessment of localized
  artificial diffusivity scheme for large-eddy simulation of compressible
  turbulent flows,} \emph{Journal of Computational Physics}, Vol. 229, No.~5,
  2010, pp. 1739--1762.

\bibitem[{Sansica(2015)}]{Sansica}
Sansica, A., \enquote{Stability and unsteadiness of transitional
  shock-wave/boundary-layer interactions in supersonic flows,} Ph.D. thesis,
  University of Southampton, October 2015.

\bibitem[{Liu et~al.(2018)Liu, Braun, and Paniagua}]{LIU2019}
Liu, Z., Braun, J., and Paniagua, G., \enquote{{Characterization of a
  Supersonic Turbine Downstream of a Rotating Detonation Combustor},}
  \emph{Journal of Engineering for Gas Turbines and Power}, Vol. 141, No.~3,
  2018.

\bibitem[{Poinsot and Lelef(1992)}]{Poinsot1992}
Poinsot, T., and Lelef, S., \enquote{Boundary conditions for direct simulations
  of compressible viscous flows,} \emph{Journal of Computational Physics}, Vol.
  101, No.~1, 1992, pp. 104 -- 129.

\bibitem[{Israeli and Orszag(1981)}]{ISRAELI1981}
Israeli, M., and Orszag, S.~A., \enquote{Approximation of radiation boundary
  conditions,} \emph{Journal of Computational Physics}, Vol.~41, No.~1, 1981,
  pp. 115 -- 135.

\bibitem[{Fang et~al.(2020)Fang, Zheltovodov, Yao, Moulinec, and
  Emerson}]{fang2020}
Fang, J., Zheltovodov, A.~A., Yao, Y., Moulinec, C., and Emerson, D.~R.,
  \enquote{On the turbulence amplification in shock-wave/turbulent boundary
  layer interaction,} \emph{Journal of Fluid Mechanics}, Vol. 897, 2020, p.
  A32.

\end{thebibliography}

\end{document}